\documentclass[twocolumn,showpacs,preprintnumbers,amsmath,amssymb,floatfix]{revtex4}

\usepackage{graphicx}
\usepackage{dcolumn}
\usepackage{bm}

\newcommand{\lsim}   {\mathrel{\mathop{\kern 0pt \rlap
  {\raise.2ex\hbox{$<$}}}
  \lower.9ex\hbox{\kern-.190em $\sim$}}}
\newcommand{\gsim}   {\mathrel{\mathop{\kern 0pt \rlap
  {\raise.2ex\hbox{$>$}}}
  \lower.9ex\hbox{\kern-.190em $\sim$}}}
\newcommand{\snoccfluxshort}{1.76^{+0.06}_{-0.05}\mbox{(stat.)}^{+0.09}_{-0.09}~\mbox{(syst.)}} 
\newcommand{\snoesfluxshort}{2.39^{+0.24}_{-0.23}\mbox{(stat.)}^{+0.12}_{-0.12}~\mbox{(syst.)}} 
\newcommand{\snoncfluxshort}{5.09^{+0.44}_{-0.43}\mbox{(stat.)}^{+0.46}_{-0.43}~\mbox{(syst.)}} 
 
\newcommand{\snomutauflux}{3.41^{+0.45}_{-0.45}\mbox{(stat.)}^{+0.48}_{-0.45}~\mbox{(syst.)}} 
\newcommand{\snoeflux}{1.76^{+0.05}_{-0.05}\mbox{(stat.)}^{+0.09}_{-0.09}~\mbox{(syst.)}} 
\newcommand{\snomutaufluxsk}{3.45^{+0.65}_{-0.62}} 
\newcommand{\snomutaufluxcomb}{3.41^{+0.66}_{-0.64}}

\newcommand{\nsigmassno}{5.3} 
\newcommand{\nsigmassk}{5.5}

\newcommand{\phisk}{2.32\pm0.03\mbox{(stat.)}^{+0.08}_{-0.07}~\mbox{(syst.)}}

\begin{document}

\title{Astrophysical Neutrino Telescopes}
\author{A. B. McDonald}
\affiliation{SNO Institute, Queen's University, Kingston, Canada K7L 3N6}

\author{C. Spiering}
 \affiliation{DESY Zeuthen, Platanenallee 6, D-15738 Zeuthen, Germany}

\author{S. Sch\"onert}
\affiliation{Max-Planck-Institut f\"ur Kernphysik, Saupfercheckweg 1, 69117 Heidelberg, Germany
}

\author{E. T. Kearns}
\affiliation{Boston Univ., Dept. of Physics, 590 Commonwealth Ave.,
Boston, MA 02215 USA}
\author{T. Kajita}
\affiliation{Institute for Cosmic Ray Research, Univ. of Tokyo, Kashiwa-no-ha 5-1-5, Kashiwa,
Chiba 277-8582, Japan}

\date{\today}

\begin{abstract}
This review describes telescopes designed to study neutrinos from astrophysical sources.
These sources include the Sun and Supernovae emitting neutrino energies up to tens of MeV, atmospheric
neutrino sources caused by cosmic ray interactions and other sources generating neutrino energies
ranging up to $1\times 10^{20}$ eV. 
Measurements with these telescopes also provide information on neutrino properties including clear
evidence for neutrino flavor change. Telescopes in operation in the past and present are described,
along with plans for future instruments to expand this rapidly growing field
of particle astrophysics.

\end{abstract}

\maketitle

\section{Introduction}

A new generation of astronomical instruments has evolved, starting with the radiochemical
 solar neutrino detector of Ray Davis and his collaborators~\cite{davis1} more than 30 years ago.
 This generation of instruments is part of the
 developing field of particle astrophysics, where particles from astrophysical objects are used
 to provide information about those objects and about the properties of the particles themselves.
 Since astrophysical objects can provide intensities and energies far outstripping any terrestrial
 source, this area of study has become a significant one for the field of particle physics. Neutrinos
 have a number of favorable properties that make them attractive as a means to study astrophysical
 objects, so neutrino telescopes have significant value for astrophysics as well.

Neutrinos have no electric charge and are not sensitive to the strong interaction. Their mass and
 magnetic moments are very small and so they undergo very few interactions even in the densest
 matter. This makes them ideal candidates for the study of otherwise inaccessible astrophysical regions,
 such as the center of the Sun, or distant astrophysical sources without distortion of their trajectories by
 intergalactic magnetic fields or material. Neutrinos are known to have three active species (electron, muon, tau)
 associated with these three lepton generations. Experiments on the decay properties of the Z Weak
 Boson \cite{PartData} have shown that there are three (2.981 $\pm$ 0.008) such active species.

The energies of neutrinos detected by the telescopes described in this article include: from a few 100 keV
 to 30 MeV for
 solar and supernova neutrinos; 0.1 GeV to a few TeV for neutrinos generated by cosmic rays interacting with the
 atmosphere; and much higher energies, ranging up to $1\times 10^{20}$ eV for the highest energy neutrino sources.
 The detection techniques differ substantially, but all detectors are situated in locations
 where interfering backgrounds are shielded to a substantial degree because the experiments tend to have very
 low neutrino detection rates as a result of the small interaction probability. The experiments seeking neutrinos
 below about 1000 GeV are typically in underground locations to range out cosmic ray background and the highest
 energy experiments are typically under ice or under water in order to combine effective shielding with very
 large detection volumes.

The sections of the paper include: 2. Solar Neutrino Detectors; 3. Supernova Neutrino Detectors;
 4. Atmospheric Neutrino Detectors; 5. Very High Energy Neutrino Detectors. This subdivision of detector groups
 is somewhat arbitrary since there is substantial overlap in the capabilities of the detectors, and all of them have
 some degree of sensitivity for supernova neutrino detection. However, this choice has some advantages
 for description, as the detectors are described in the section covering the primary scientific motivation
 for the detector and reference to additional capabilities may be mentioned further in another section.
 This paper is mainly intended to describe the principal features of the detectors themselves, but in each case
 the physics motivation for the detectors is presented, along with the detector scientific capability and a
 succinct summary of the results obtained to date or anticipated for the future.

There are many astrophysical motivations for the detection of neutrinos with neutrino telescopes, as well as
 opportunities to determine particle physics properties of neutrinos themselves through the use of astrophysical
 sources. The complementary nature of this information has made neutrino telescopes a central part of the rapidly
 developing field of Particle Astrophysics.

Some of the principal astrophysical motivations are as follows. The neutrinos produced in the Sun arise from the
 principal nuclear reactions powering the Sun. Detailed models of the Sun have been developed that match other
 measured solar properties very well and predict a spectrum of neutrinos from a number of the nuclear reactions.
The study of neutrinos from the Sun over the full range of energies provides a very detailed test of solar models,
 including otherwise inaccessible information from the core region. 

Atmospheric neutrinos arise primarily from
 the decay of muons, pions or kaons produced by energetic cosmic rays interacting with the atmosphere.
 Measurements of neutrino fluxes over a broad range of energies can be combined with other sources of
 information to obtain an understanding of cosmic rays from astrophysical sources. The highest
 energy neutrinos arise from high energy cosmic rays, protons or nuclei. One of the objectives of neutrino
 detection is to determine the origins and acceleration mechanisms of these very high energy charged particles.
 Neutrinos from Supernovae include electron neutrinos arising from the conversion of neutrons to protons during
 the initial infall as well as electron, mu and tau neutrinos and anti-neutrinos produced during the expansion
 phase following the point of maximum compression. Neutrinos carry away the majority of the energy from
 a supernova and can provide very detailed information on the processes involved in these spectacular
 astronomical events. 

Neutrinos from astrophysical sources can also be used to study detailed properties of neutrinos and results
 from these measurements have provided very valuable information, including the first clear indications of
 neutrino flavor change (i.e. the change of a neutrino of one family into one of another family) and
 information on neutrino mass differences. Some of the particle physics information
 that has been obtained from astrophysical neutrinos is as follows.

In the case of neutrino mixing,
the neutrino flavor
fields
$\nu_{{\ell}}$
are superpositions
of
the components
$\nu_{k}$
of the fields of neutrinos with definite masses
$m_k$:
\begin{equation}
\nu_{{\ell}}
=
\sum_{k=1}^{3}
U_{{\ell}k}
\,
\nu_{k}
\qquad
(\ell=e,\mu,\tau)
\,,
\label{mixing}
\end{equation}

\noindent
where $U$ is the 3 x 3 unitary Maki-Nakagawa-Sakata-Pontecorvo (MNSP) mixing matrix \cite{MNSP}. 

When the neutrinos travel in a vacuum or low density region, the linear combinations
 are changed so that the detected neutrino types appear to "oscillate" among the neutrino
 flavors. If the masses are not strongly degenerate, then the 3 x 3 MNSP matrix can be
 separated into 2 x 2 matrices affecting oscillation among two masses
 almost independently of the third. 
In this case
there is one $\Delta m^2$ such as $m_2^2-m_1^2$ for the masses involved and the mixing matrix
\begin{equation}\label{vacangle}
U = \left( \begin{array}{rr} \cos \theta & \sin \theta \\
-\sin \theta & \cos \theta \end{array} \right)
\end{equation}

For this simplest case where two mass eigenstates dominate the
 process, the following probability is predicted for subsequent detection of a given
 neutrino type after it has traveled for a distance L in vacuum:

\begin{displaymath}
P = 1 - sin^2(2\theta) sin^2(1.27 \Delta m^2 L/E)
\end{displaymath}
\noindent
where $\Delta m^2$ is in eV$^{2}$, $L$ is the source-detector distance in meters, $E$ is
the neutrino energy in MeV and $\theta$ is a mixing angle.
In order for a significant fraction of neutrinos to change their flavor, $sin^2 (2 \theta)$
 must be reasonably large and $\Delta m^2 L/E$ must be appropriate for the source and detector
 involved. This process has been used to interpret atmospheric neutrino data observed
 by the Super-Kamiokande detector as described in Section 4 wherein muon neutrinos
 appear to oscillate predominantly to tau neutrinos as they traverse the Earth. These
 measurements can be used to define elements of the MNSP matrix, in particular, $\Delta m_{23}^2$ 
and $sin^2 \theta_{23}$.

Additional effects can occur when neutrinos pass through regions of dense matter.
 These effects are associated with the fact that electron neutrinos can have interactions
 with electrons through W exchange in addition to the Z exchange occurring for all
 flavors. These effects can result in "matter enhancement" of the oscillation process \cite{MSW},
 the so-called MSW (Mikheyev-Smirnov-Wolfenstein) effect. This enhancement greatly broadens the region
 of $\Delta m^2_{12}$ and $sin^2 \theta_{12}$ where significant effects can occur and introduces an additional dependence
 on the neutrino energy. 

The Sudbury Neutrino Observatory (SNO) has provided evidence \cite{SNO} for neutrino flavor change
 by observing clear differences between one reaction sensitive to only electron neutrinos
 and another sensitive equally to all neutrino types. The results show that about 2/3 of
 the electron neutrinos produced in the Sun appear to have transformed to other types
 before detection on Earth. A similar conclusion is reached by comparing the SNO results
 with the measurements of solar neutrinos by Super-Kamiokande \cite{SK}, using a reaction with a
 small sensitivity to other neutrino types in addition to electron neutrinos. These
 results, together with a number of other measurements of solar neutrinos with different
 energy thresholds, can be interpreted in terms of neutrino oscillations, providing
 additional information about elements of the MNSP matrix, particularly
 $\Delta m_{12}^2$ and $sin^2\theta_{12}$.

Many other particle physics properties of neutrinos can be defined by measurements
 with neutrino telescopes, including limits on neutrino decay lifetimes and magnetic
 moments. Future experiments will also improve significantly on present neutrino mass
 and mixing parameters, particularly when combined with measurements from terrestrial
 sources such as reactors and accelerators.

The improved definition of neutrino properties through these measurements from
 astrophysical sources can, in turn, be used to improve our astrophysical information.
 For example, with the oscillation properties of neutrinos defined, it is possible
 to understand the original fluxes of solar electron neutrinos as a function of energy.
 This provides a very accurate test of solar models and adds to our understanding
 of the astrophysics of the Sun. 

\section{Solar Neutrinos}

\subsection{Physics and Astrophysics Motivations}

\subsubsection{Astrophysics Motivations}

Solar models have been developed to a high level of sophistication \cite{SSM1,SSM2} and are
 capable of correct predictions for many of the measurable solar properties, including
 recent helioseismological results that probe the inner regions with unprecedented accuracy.
The detection of neutrinos from the Sun provides information on the nuclear
 reactions and associated neutrino fluxes that pin down the properties of the solar core
 region with considerable accuracy. Neutrino fluxes have a strong dependence on the
 temperature of the core region and on other processes such as mixing of materials between various
 solar regions. Models of the Sun following standard assumptions \cite{SSM1,SSM2}
 are found to be remarkably successful for all properties other than neutrino fluxes.
However, initial neutrino
 measurements either solely or primarily sensitive to electron neutrinos were found to be lower
 than expected because of the presence of neutrino flavor-changing processes.
 The possibility of a finite neutrino magnetic moment also enables solar magnetic fields
 to be probed with sensitivities restricted by current limits on the magnetic moments set by
 terrestrial or other astrophysical determinations.

\begin{figure}
\includegraphics[height=9cm,angle=270]{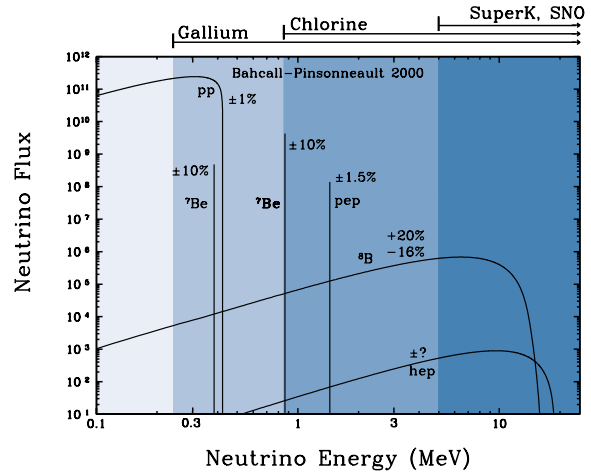}
        \caption{\label{neutrinospectrum}Spectrum of neutrinos from the Sun as obtained by a Standard Solar Model calculation}
\end{figure}

The set of nuclear reactions thought to be primarily responsible for energy generation in
 the Sun are shown in Table 1 and fluxes predicted by a Standard Solar Model Calculation \cite{SSM1} 
assuming no neutrino flavor change are shown
 in Figure~\ref{neutrinospectrum}.

\begin{table*}
\caption{\label{tab:SSMnumbers} Principal nuclear reactions producing neutrinos in the Sun \cite{SSM1}}
\begin{ruledtabular}
\begin{tabular}{lllll}
Neutrino Source &Energy & Flux at Earth  & $^{35}$Cl Production & $^{71}$Ga Production  \\
& [MeV] & [$10^{10}$ cm$^{-2}$ s$^{-1}$] & [SNU] & [SNU] \\
\hline 
pp ($p p \rightarrow d  e^+ \nu_e$) & $< 0.42$ & 6.0 & ---- & 69.7 \\
pep ($p e^- p \rightarrow d \nu_e$) & 1.4 & 0.014 & 0.22 & 2.8 \\
$^{7}$Be ($^{7}$Be e$^{-} \rightarrow ^{7}$Li $\nu_e$) & 0.38, 0.86  & 0.48 & 1.2 & 34.2 \\
$^{8}$B ($^{8}$B $\rightarrow ^{8}$Be$^{*}$ e$^{+} \nu_e$) & $< 15$ & 0.00051  & 5.8 & 12.1 \\
$^{3}$He p ($^{3}$He p $\rightarrow ^{4}$He e$^{+}$ $\nu_e$) & $< 18.77$ & $9.3 10^{-7}$  & 0.04 & 0.1 \\
$^{13}$N & $< 1.20$ & 0.055  & 0.09 & 3.4 \\
$^{15}$O & $< 1.73$ & 0.048 & 0.33 & 5.5 \\
$^{17}$F & $< 1.74$  & 0.00056  & 0.0 & 0.1 \\
Total &  & 6.5 & 7.6 & 128 \\
\end{tabular}
\end{ruledtabular}
\end{table*}

\subsubsection{Particle Physics Motivations}

As discussed above, the possibility to use copious quantities of neutrinos from the Sun to study the basic properties
 of neutrinos provides a further motivation for such measurements. The long baseline from the Sun to the Earth provides
 an opportunity for neutrinos to change their flavor through oscillation processes and the high density of electrons in the
 Sun can provide additional enhancement of such flavor change \cite{MSW}. Limits on neutrino decay and on neutrino
 magnetic moments can also be obtained from such studies.

\subsection{Solar neutrino detection reactions}

Solar neutrino detection reactions that have been used successfully to date fall into two main categories.
 1. Radiochemical measurements involve the transformation of atoms of an element such as Cl or Ga into another
 radioactive element through inverse
 beta decay induced by electron neutrinos. These reactions can be observed by sweeping
 the radioactive elements from the detector volume and observing the subsequent decay.  The measurements typically
 involve collection periods on the order of a month and subsequent decay measurement periods of many months.
 2. Real-Time measurements involve the observation of events produced by neutrino interactions in the detector medium
 in real time. Experiments to date have used light and heavy water as media for real-time neutrino detection.
 The reactions
 that have been used involve elastic scattering of neutrinos from electrons, the inverse beta decay reaction 
on deuterium
 producing energetic electrons and inelastic scattering on deuterium producing free neutrons. 

\subsubsection{Cherenkov Detection Process}

The real time detection reactions (including the capture of the neutrons produced in the latter reaction) all
result in energetic electrons being produced that are observed via the Cerenkov process. Charged particles travelling
in a transparent medium at greater than the group velocity of light in that medium radiate Cherenkov light. These photons
have a continuous spectral distribution and are emitted in a forward cone whose axis is the direction of the electron and
whose opening angle ($\theta$) is given by

\begin{displaymath}
cos ~\theta = (n \beta)^{-1}
\end{displaymath}

\noindent
where $n$ is the index of refraction in the medium and $\beta$ is the electron's speed relative to the speed of light.
For light and heavy water $\theta$ equals about 41 degrees for relativistic electrons. The number of photons emitted
by the electrons is approximately proportional to the track length (and hence energy) and is about 350 photons per cm
in the typical spectral range for which photomultipliers are sensitive (about 300 to 600 nanometers). The electron
track length is about 0.45 cm per MeV in water for kinetic energies between 5 and 15 MeV, thus about 1100 photons are produced
by a 7 MeV electron.

\subsection{Radiochemical Detectors}
\subsubsection{Homestake Cl Experiment}

The first detector to report fluxes of neutrinos from the Sun was the pioneering experiment of Ray Davis and
 co-workers \cite{davis1} using 680 tons of liquid perchlorethylene sited 1480 meters underground (4300 meters water equivalent)
in the Homestake
 gold mine near Lead, South Dakota, USA. This experiment employs the electron neutrino capture reaction on $^{37}$Cl,
 producing $^{37}$Ar atoms in the liquid. The radioactive $^{37}$Ar atoms (35-day half life) are swept from the
 containment tank
 by flushing with helium gas about every 100 days. The $^{37}$Ar atoms are condensed into sensitive proportional counters
 with high efficiency and the resulting decays are observed for many half-lives of $^{37}$Ar. Production rates of about 0.5
 argon atoms per day are observed, about three times lower than predictions of the Standard Solar Models \cite{SSM1,SSM2}.

\begin{figure}
\centering
\includegraphics[width=9cm]{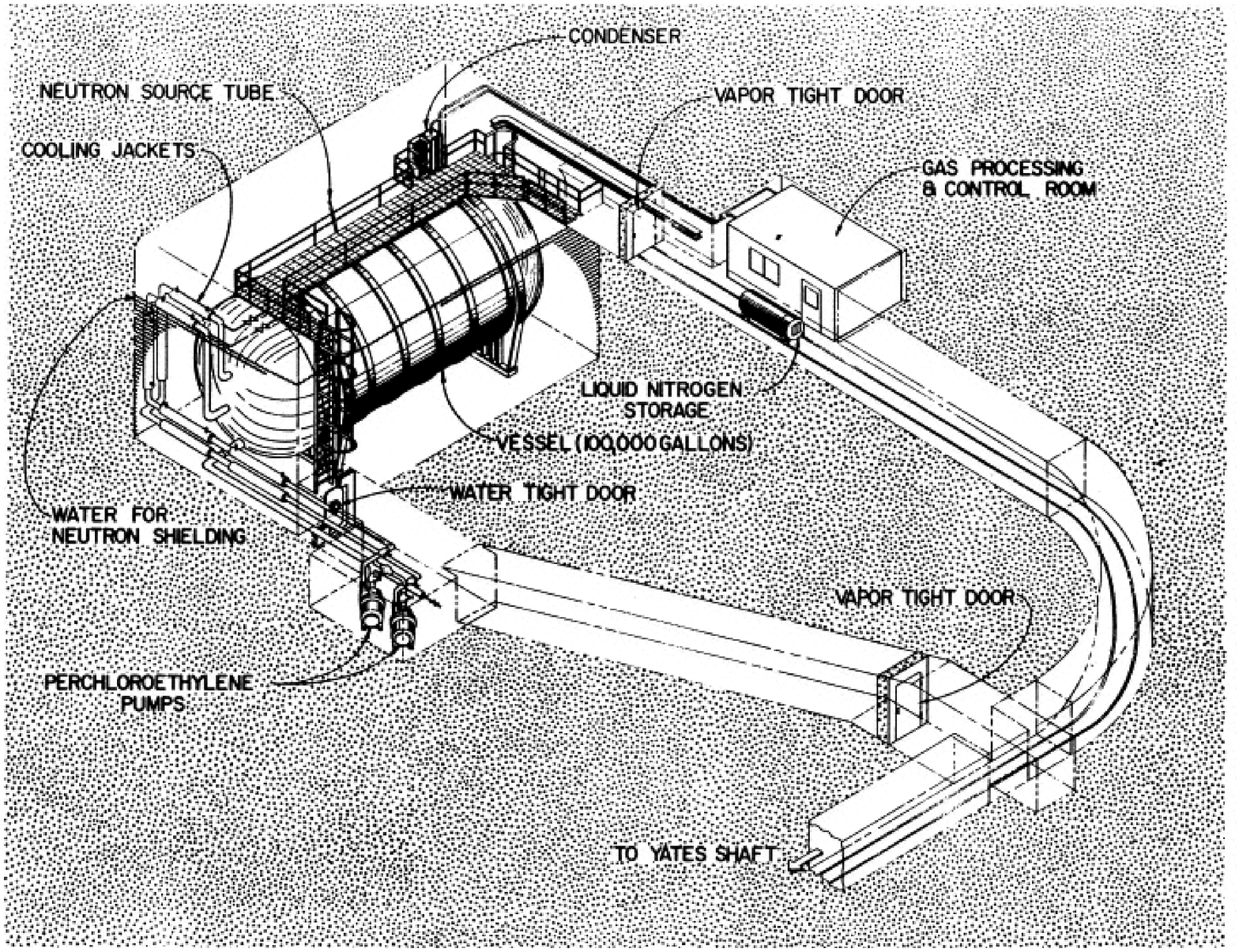}
       \caption{\label{HomestakeFigure} The Chlorine detector in the Homestake mine.}
\end{figure}

The experimental apparatus is shown in Figure~\ref{HomestakeFigure}.

 The detector is sited underground to reduce $^{37}$Ar production by cosmic rays. Detector and proportional counter
 materials were carefully selected to minimize contributions from background radioactivity. Background rates of $^{37}$Ar
 from sources other than the Sun were restricted to about 0.02 atoms per day. Non-radioactive tracer isotopes $^{36}$Ar
 and $^{38}$Ar are injected during successive runs to determine the argon extraction efficiency and are measured by mass
 spectrometry. Pulse shape analysis is used on the pulses from the proportional counter to reduce counter backgrounds
 from other than $^{37}$Ar decay.

The detector has been operated from 1968 until 2002 and the cumulative average of the data \cite{davis1} is
 $2.56 \pm 0.16(stat) \pm 0.16(syst)$ SNU, where a Solar Neutrino Unit (SNU) is defined as one electron neutrino capture
 per $10^{+36}$ atoms of $^{37}$Cl per second. This result is substantially smaller than the predictions of Standard
 Solar Models,
 for which the latest prediction \cite{SSM1} is $7.6 ^{+1.3}_{-1.1}$. The capture reaction on $^{37}$Cl is calculated to be
 sensitive primarily
 to neutrinos from $^8$B decay in the Sun, but there is substantial sensitivity to neutrinos from $^7$Be as well.
 The expected contributions from the various SSM calculated fluxes are shown in Table~\ref{tab:SSMnumbers} . 

\subsubsection{Ga-Based Experiments: SAGE, GALLEX, GNO}

The electron neutrino capture reaction on Ga ($^{71}$Ga($\nu_e$,e$^-$)$^{71}$Ge) has a threshold of 232 keV, well below the maximum energy of the
 neutrinos from the pp reaction in the Sun. Therefore detectors based on this medium have significant sensitivity to pp
 neutrinos as can be seen in Table~\ref{tab:SSMnumbers}. Experiments with Ga as the detector material are being carried
 out in the Baksan
 laboratory in Russia (SAGE) \cite{SAGE} with about 60 tons of liquid Ga metal and in the Gran Sasso laboratory in Italy
 (GALLEX, GNO) \cite{GNO} with 30.3 tons of Ga in a concentrated GaCl$_{3}$-HCl solution. The underground depths correspond
 to about 4700 meters of water equivalent for the SAGE experiment and 3300 meters for GALLEX and GNO.
 
The technique for detection of the neutrino reaction involves the observation of the radioactive decay of $^{71}$Ge
 (half-life 11.4 days) by extraction and deposition into low-background proportional counters. For SAGE, the Ge
 is removed by extraction into an aqueous solution via an oxidation reaction, followed by concentration, conversion
 to GeCl$_{4}$ and synthesis to GeH$_{4}$ for use in the proportional counter. For GALLEX and GNO, volatile
 $^{71}$GeCl$_{4}$ is formed
 directly in the target solution, swept out by nitrogen gas, absorbed in water and then converted to GeH$_{4}$ before
 insertion into the proportional counter. The K and L electron capture decay of $^{71}$Ge is observed through the
 energy deposition from the Auger electron and X-rays emitted in the decay. Detector materials are very carefully
 selected to minimize background counting in the proportional counter. Pulse shape discrimination is used to
 distinguish
 the highly-localized pulses from electron capture from the spatially-extended tracks from higher energy electrons
 associated with background processes.

Non-radioactive isotopes of Ge are inserted into the detector media and extracted along with the $^{71}$Ge to monitor
 extraction efficiency. Analysis of the counting rates includes fitting a component with the characteristic half-life of $^{71}$Ge
 as well
 as for background components including cosmogenically-produced $^{68}$Ge (half-life 271 days) and other long-lived
 backgrounds. Special precautions are taken in each of the experiments to reduce contributions from $^{222}$Rn (3.8 days)
 to a very small level. The experiments have used intense $^{51}$Cr sources \cite{Cr51} to calibrate directly the sensitivity of the
 detector to neutrinos.

The counting rates from these two measurements to date \cite{SAGE,GNO} are as follows:
GALLEX + GNO: $70.8 \pm 4.5 (statistical) \pm 3.8 (systematic)$ SNU, SAGE: $70.9 +5.3/-5.2 (stat) +3.7/-3.2 (syst)$ SNU. 
These numbers are in excellent agreement and are much smaller than the predictions of the standard solar model
 (132 SNU) as shown in Table~\ref{tab:SSMnumbers}.

\subsection{Real-Time Solar Neutrino Detectors}

\subsubsection{Kamiokande and Super-Kamiokande}

As described in Section 3, the Kamiokande detector, 1000 meters underground (2600 meters water equivalent)
in the Kamioka mine in Japan, started
 operation in 1983 with the primary objective to study proton decay in a large light water volume. The detector had
 a fiducial volume of about 1 kton for the study of high energy events, including events produced by atmospheric neutrinos.
 After several years of operation, it was decided to add the additional capability of detecting solar neutrinos by
 improvements to the detector systems, particularly the electronics systems and the anti-counter system with full
 solid angle coverage. The detection reaction was elastic scattering
 of neutrinos from electrons, observed by the detection of Cherenkov light from the recoiling electron as discussed
 in Section 3. For the much lower energy solar neutrino measurements, the central region of light water (680 tons) was used
 and an energy threshold of about 7.0 MeV was achieved, limited primarily by gamma rays from residual radioactivity
 in the water, particularly $^{222}$Rn. With this threshold, the solar neutrino sensitivity of the detector
 was almost exclusively for neutrinos
 from $^8$B decay, with the neutrinos from the $^3$Hep reaction expected to make a tiny contribution. A set of careful
 calibrations were carried out, including the use of radioactive sources of gamma rays. 

The elastic scattering reaction has a strong directional dependence, peaked away from the direction of the incoming
 neutrino. A substantial excess of events from the Sun was observed, clearly distinguishable from background events
 and confirming the solar origin of the neutrinos. The flux of neutrinos from the $^8$B decay was deduced \cite{KAMIOKA} to be
 $2.8 \pm 0.19 (stat) \pm 0.33 (syst) \times 10^{+6}$ cm$^{-2}$ s$^{-1}$, clearly less than the solar model
 prediction \cite{SSM1} of $5.05 \times 10^{+6}$ cm$^{-2}$ s$^{-1}$.
 There was no
 evidence for variation in the flux as a function of time within the uncertainties.

\begin{figure}
\includegraphics[width=3.1in]{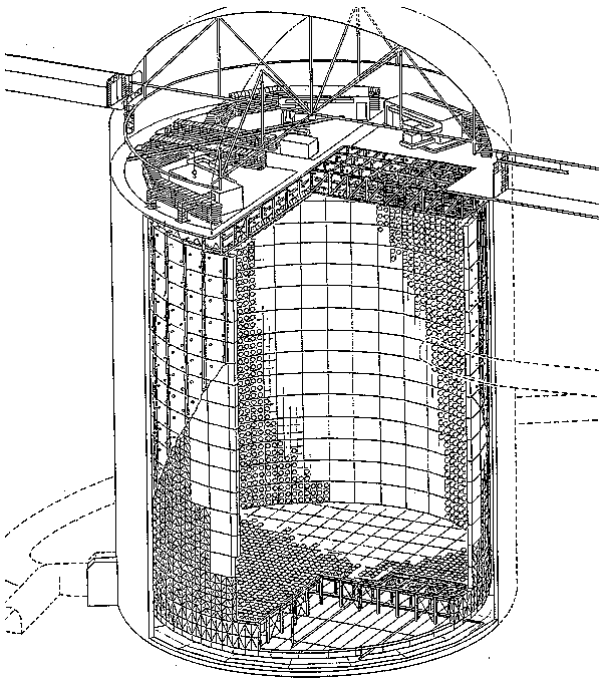}
\caption{\label{fig:skdet}
The Super-Kamiokande detector.}
\end{figure}

The success of the Kamiokande detector prompted the construction of a much larger Super-Kamiokande detector
 nearby in the same mine as shown in Figure~\ref{fig:skdet}. After completion of the Kamiokande experiment,
 the detector was dismounted, the cavern enlarged  and a new liquid scintillator experiment, KamLAND, with
 1 kton target mass constructed. The objective of KamLAND is the detection of electron anti-neutrinos (${\bar \nu}_e$) from distant
 nuclear power reactors to probe the solar large mixing angle parameter space. Data taking commenced in early 2002 and
 first results have been published \cite{KAMLAND}.

\begin{figure}
\includegraphics[width=3.1in]{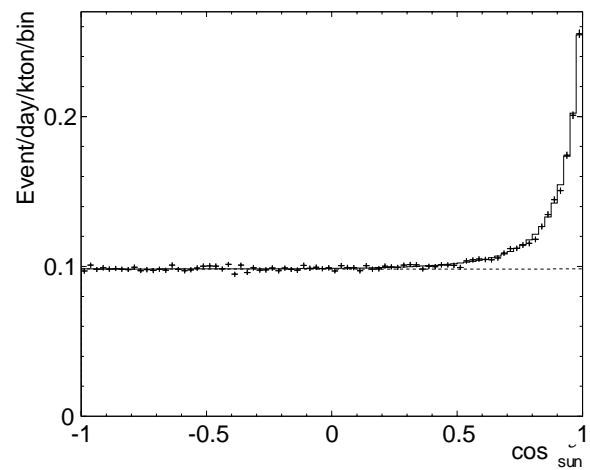}
\caption{\label{fig:skcostheta}
Angular distribution of events from the Super-Kamiokande detector with respect to direction from the Sun.}
\end{figure}

\begin{figure}
\includegraphics[width=3.1in]{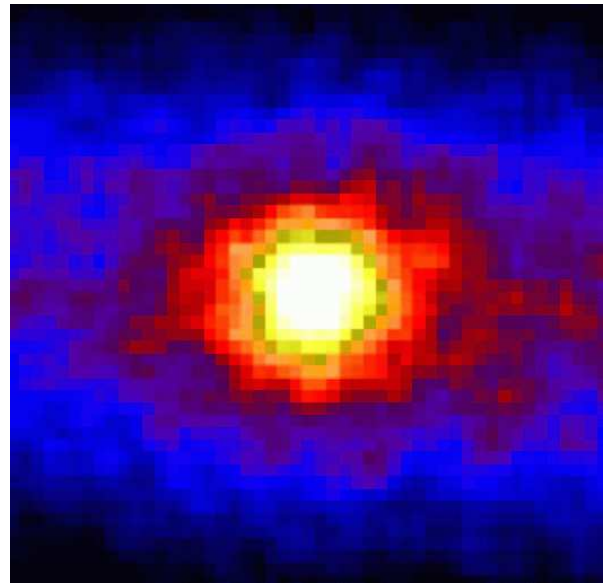}
\caption{\label{fig:skymap}
Neutrino profile of the Sun obtained by the Super-Kamiokande detector. One pixel corresponds to one degree, which may
 be compared to the $1/2$-degree apparent size of the sun. The distribution of colors is determined by the angle
 of neutrino-electron scattering.}
\end{figure}

A detailed description of the Super-Kamiokande detector is given in the section on atmospheric neutrinos below.
For solar neutrino detection via the elastic scattering of neutrinos, the fiducial volume in the Super-Kamiokande
 detector is 22.5 kton. Considerable care was taken
 to reduce the radioactive backgrounds in the detector during construction. With this improved detector, the counting
 rate was increased substantially and additional improvements were made in calibration through the use of an electron
 linac. Figure~\ref{fig:skcostheta} shows the data obtained for 1496 days of counting \cite{SK} versus the direction
 from the Sun for
 a threshold of about 5 MeV. The excess events from the Sun are clearly observable. Figure~\ref{fig:skymap} shows a map
 of the Sun obtained with neutrinos detected by Super-Kamiokande.

 The measured solar neutrino
 flux from $^8$B decay is $2.35 \pm 0.02 (stat) \pm 0.08 (syst) \times 10^{+6}$ cm$^{-2}$ s$^{-1}$, also significantly
 smaller than the solar model prediction. The neutrino energy spectrum is very similar to that expected from $^8$B decay.

\begin{figure}
\includegraphics[width=3.1in]{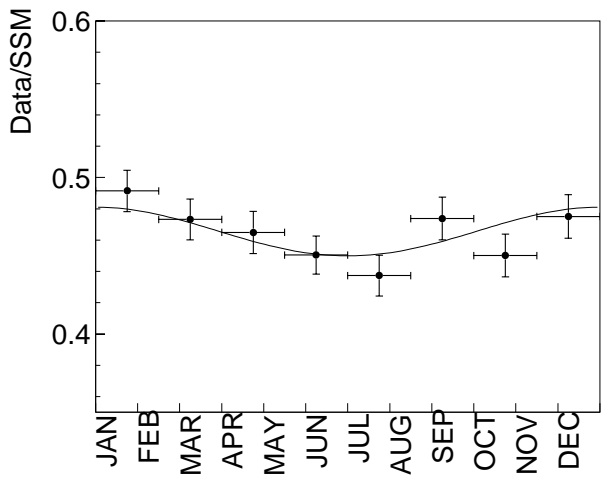}
\caption{\label{fig:skseasonal}
Seasonal variation of the solar neutrino flux. The curve shows the expected seasonal variation of the flux introduced
by the eccentricity of the Earth's orbit.}
\end{figure}

If there is a significant neutrino magnetic moment, this could couple with solar magnetic field distributions
 and produce different observed fluxes through flavor change as neutrinos pass through different solar regions at
 various times of year~\cite{DayNight}. However, the neutrino flux as a function of time of year is observed to follow the variation
 expected for the Earth-Sun distance variation with no observable additional effects as shown in Figure~\ref{fig:skseasonal}.
 Studies over longer periods of time to seek correlations between the data and the sunspot numbers (indicative of magnetic
 fields in the outer regions of the Sun) also show no correlations~\cite{SK}.

Variation of the observed neutrino flux was also studied as a function of the zenith angle of events reaching the detector
 to search for effects of matter enhancement of flavor change while the neutrinos pass through the Earth. No clear 
effect was observed~\cite{SK} as can be seen from the measurement of the fractional difference in the average rate in the day
 compared to the night: A = $ -0.021 \pm 0.020 (stat) ^{+0.013}_{-0.012} (syst.)$.

To this point, the data from the radiochemical and the light water detectors all showed a significant reduction in the
 detected neutrino flux compared to the standard solar models. Many studies were made of solar models to seek
 changes that could produce neutrino fluxes similar to those observed. As no obvious changes could be found to
 explain the neutrino measurements there was a strong indication that neutrino oscillations could be occurring. However,
 since the different experiments were sensitive to different combinations of fluxes from the solar reactions, analyses
 in terms of neutrino oscillations depended on the correctness of solar models.

\subsubsection{Sudbury Neutrino Observatory}

\begin{figure}
\includegraphics[width=3.1in]{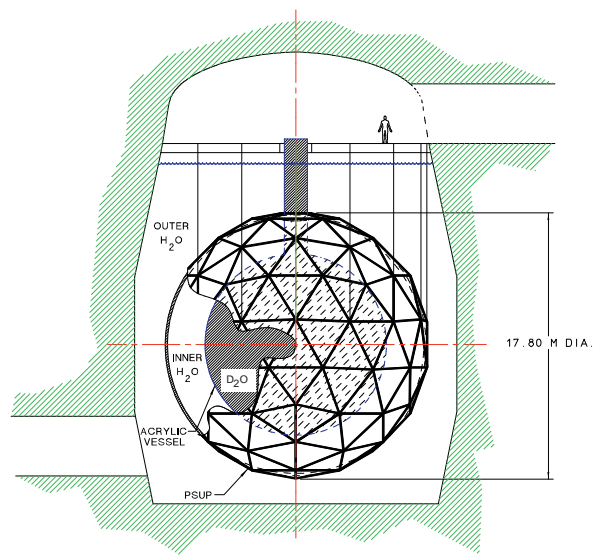}
\caption{\label{fig:snooutline}
Outline drawing of the SNO detector.}
\end{figure}

The Sudbury Neutrino Observatory~\cite{SNONIM} is a real-time neutrino detector based on 1100 tons of heavy water
 sited 2000 meters underground (6200 meters water equivalent) near Sudbury, Ontario, Canada. Neutrinos are detected
 via the Cherenkov light produced by
 electrons moving faster than the speed of light in the central heavy water volume or in a surrounding volume
 of light water.
 As shown in Figure~\ref{fig:snooutline}, the heavy water is contained in a 12 meter diameter transparent acrylic
 plastic vessel (AV) and the
 Cherenkov light is detected by 9456 20-cm photomultiplier tubes mounted on a 17.8 meter geodesic sphere (PSUP). The
 detector is contained within a polyurethane-lined cavity 22 meters in diameter by 34 meters high, filled with
 ultra-pure light water.

SNO measures the $^8$B neutrinos from the Sun through the reactions:
 \begin{center}
  \begin{tabular}{ll}
     $\nu_e + d \rightarrow p + p + e^-$\hspace{0.5in} & (Charged Current:CC)\\
     $ \nu_x + d \rightarrow p + n + \nu_x$ & (Neutral Current:NC)\\
     $ \nu_x + e^- \rightarrow \nu_x + e^-$  & (Elastic Scattering:ES)\\        
  \end{tabular}
 \end{center}

The experimental objective is to use this combination of reactions to determine if electron neutrinos are changing
 their flavor, without reference to solar model calculations. The CC reaction is sensitive exclusively
 to electron neutrinos,
 while the NC reaction is sensitive to all neutrino flavors ($x = e, \mu, \tau$) above the energy threshold of
 2.2 MeV. The ES
 reaction is sensitive to all flavors as well, but with reduced sensitivity to $\nu_{\mu}$ and $\nu_{\tau}$.
 Comparison of the $^8$B flux deduced
 from the CC  reaction with the total flux of active neutrinos observed with the NC reaction (or the ES reaction with
 reduced sensitivity) can provide clear evidence of neutrino flavor transformation without reference to solar model
 calculations. If the neutrino flavor change process is occurring and can be fully understood, the original
 flux of electron
 neutrinos from the Sun can also be determined, providing an excellent test of solar model calculations. 

The CC and ES reactions are observed through the Cherenkov light produced by the electrons. The electronic systems
 are designed to obtain a measure of the time of arrival and
 light intensity observed by each of the PMT's. The two dimensional projection of the Cherenkov cone on the array
 of PMT's and the arrival times provide a measure of the light intensity, origin and direction of the event. The NC reaction is observed
 through the detection of the neutron in the final state of the reaction.  The SNO experimental plan involves
 three phases
 wherein different techniques are employed for the detection of neutrons from the NC reaction. During the first phase,
 with pure heavy water, neutrons are observed through the Cherenkov light produced when neutrons are captured on
 deuterium, producing 6.25 MeV gammas. For the second phase, about 2.7 tons of salt is added to the heavy water
 and neutron detection is enhanced through capture on Cl, with about 8.6 MeV gamma energy release and higher
 capture efficiency. In the third phase, the salt is removed and an array of $^{3}$He-filled proportional counters
 is installed
 to provide neutron detection independent of the PMT array.

The SNO detector was constructed from carefully chosen, low-radioactivity materials. All workers took showers and
 used lint-free clothing to maintain Class 2000 air quality during construction. The water systems include
 processes for
 purifying the light and heavy water and measuring the residual radioactivity accurately. During operation,
 radioactivity
 levels were achieved that result in background events for the NC reaction that are less than 5\% of the predicted rate
 from the Standard Solar Model neutrino flux. The threshold for observation of the CC and ES reactions is 5 MeV in 
equivalent electron energy.

\begin{figure}
\includegraphics[width=3.1in]{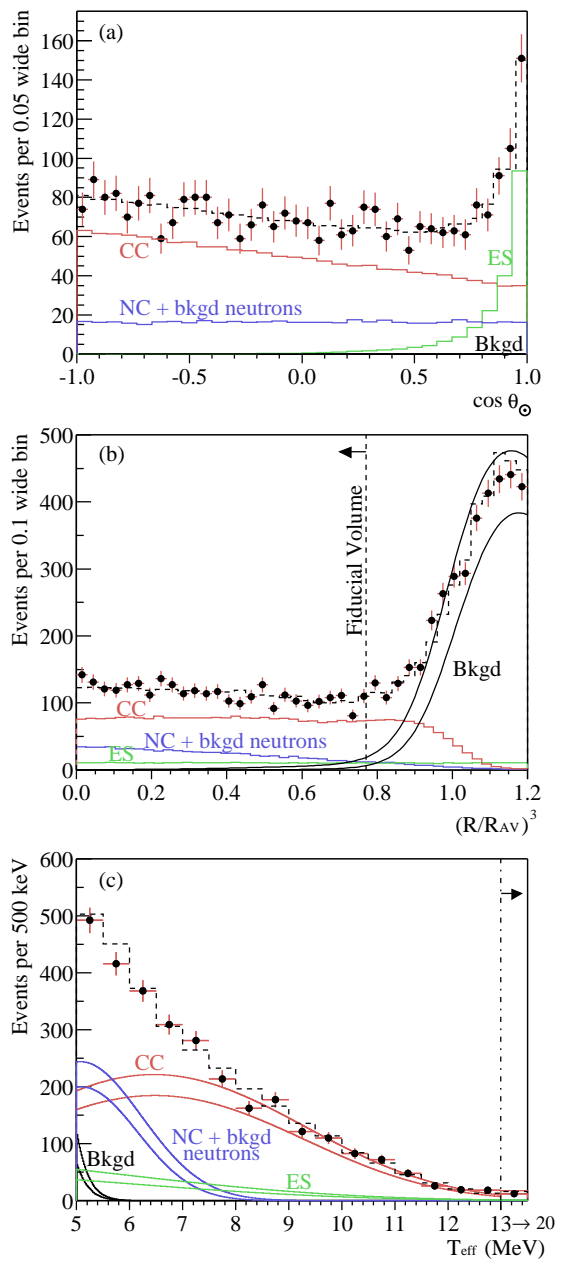}
\caption{\label{fig:snodata}
(a) Distribution of $\cos\theta_{\odot}$ for $R \le 550$ cm.
 (b) Distribution of the volume weighted radial variable $(R/R_{\rm AV})^{3}$.
  (c) Kinetic energy for $R \le 550$ cm.  Also shown are the Monte Carlo predictions for CC, ES
 and NC + bkgd neutron events scaled to the fit results, and the calculated spectrum of Cherenkov
 background (Bkgd) events.
  The dashed lines represent the summed components, and the bands show $\pm 1\sigma$ uncertainties.}
\end{figure}

Results from the initial phase of the experiment with pure heavy water~\cite{SNO1,SNO,SNO3} show clear evidence for
 neutrino flavor change,
 without reference to solar model calculations. The data from this phase is shown in Figure~\ref{fig:snodata}, together with
 the best fit
 to the data of the NC (6.25 MeV gamma) shape, the CC and ES reactions assuming an undistorted shape for the $^8$B
 neutrino spectrum and a small background component determined from independent measurements.

The fluxes inferred for this fit are:
\begin{eqnarray*}
\phi^{\text{SNO}}_{\text{CC}} & = & \snoccfluxshort \\
\phi^{\text{SNO}}_{\text{ES}} & = & \snoesfluxshort \\
\phi^{\text{SNO}}_{\text{NC}} & = & \snoncfluxshort. 
\end{eqnarray*}
\noindent
where these fluxes and those following in this section are quoted in units of 10$^{6}$ cm$^{-2}$ sec$^{-1}$.
A simple change of variables resolves the data directly into electron ($\phi_{e}$) and non-electron
($\phi_{\mu\tau}$) components. This change of variables allows a direct test of the null
hypothesis of no flavor transformation ($\phi_{\mu\tau}=0$) without requiring calculation of the
CC, ES, and NC signal correlations.
  
\begin{eqnarray*}
\phi_{e} & = & \snoeflux \\
\phi_{\mu\tau} & = & \snomutauflux 
\end{eqnarray*}
\noindent assuming the standard ${}^{8}$B shape.

Combining the statistical and systematic uncertainties in quadrature,
 $\phi_{\mu\tau}$ is $\snomutaufluxcomb$,
which is \nsigmassno$\sigma$  above zero, providing  strong
evidence for flavor transformation consistent with neutrino oscillations~\cite{MNSP}.
  Adding the Super-Kamiokande ES measurement of the ${}^{8}$B flux~\cite{SK} 
$\phi^{\text{SK}}_{\text{ES}}=\phisk$ as an additional constraint,
 we find $\phi_{\mu\tau}=\snomutaufluxsk$, which is \nsigmassk$\sigma$ above zero.

The total flux of ${}^{8}$B neutrinos measured with the NC reaction is in very good agreement with
the SSM prediction~\cite{SSM1} of $5.05 \pm 1.0 \times 10^6$ cm$^{-2}$ s$^{-1}$. 

(Note added in proof: The SNO collaboration has recently reported~\cite{SNOsalt} results from the second phase of the
 experiment where sodium chloride was added to the heavy water. The neutrons from the NC reaction capture on Cl
and provide a more isotropic pattern at the PMT's for these events than for the CC events. This enables the fluxes
 from the CC and NC reactions to be determined independently with no constraint on the shape of
the CC energy spectrum. The results are in good agreement with the previous constrained measurements and the total flux
 of ${}^{8}$B neutrinos inferred from the NC reaction was determined to be
 $5.21 \pm 0.27~\mbox{(stat)}~\pm0.38~\mbox{(syst)} \times 10^6$ cm$^{-2}$ s$^{-1}$. This determination of the
${}^{8}$B flux is independent of the presence of oscillations to active neutrino types and is in very good
 agreement with SSM predictions.)

\subsection{Summary of Results from Solar neutrino telescopes to date}
\begin{figure}
\includegraphics[width=3.1in]{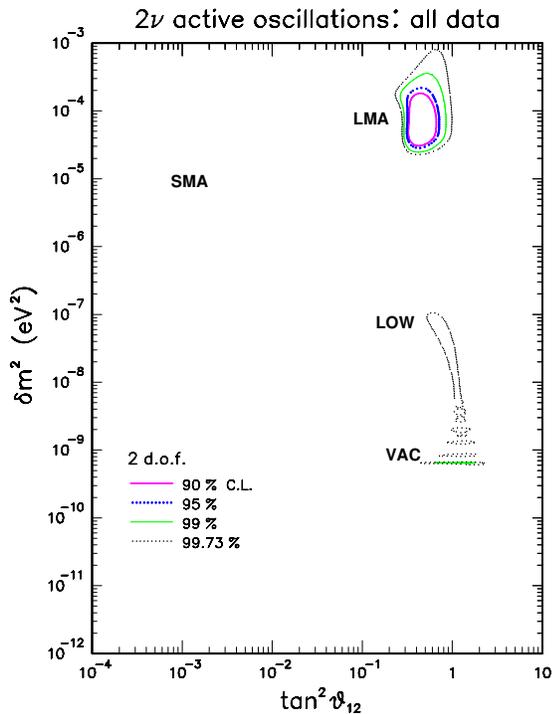}
\caption{\label{fig:LisiMSW}
Global fit to all the solar neutrino data in terms of a two-neutrino MSW oscillation scenario~\cite{Fogli}}
\end{figure}

\begin{figure}
\includegraphics[width=3.1in]{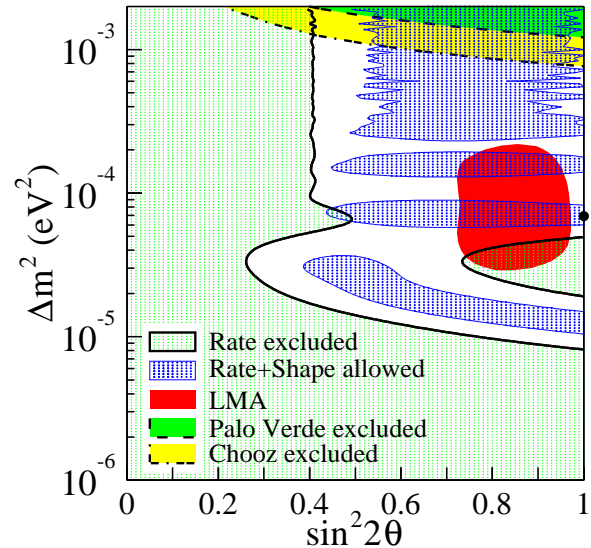}
\caption{\label{fig:KamlandExcPlot}
Allowed regions of anti-neutrino oscillation parameters for the rate analysis and the combined rate and shape
 analysis from KamLAND at 95\% confidence limit. Also shown are the LMA region allowed by solar neutrino
 measurements and excluded regions from CHOOZ and Palo Verde reactor neutrino experiments.}
\end{figure}

The measurements to date (see Table~\ref{tab:solardets}) have provided convincing evidence for flavor change of neutrinos and matter enhancement
 of these effects in the Sun. If all the measurements are combined in a global fit~\cite{Fogli} to the neutrino oscillation
 process
 (assuming 2 neutrino oscillation), the remaining allowed regions are shown in Figure~\ref{fig:LisiMSW}. These regions
 are labeled
 LMA, LOW and VAC, and the most favored among them is the LMA region. Recent measurements~\cite{KAMLAND} reported
 by the KamLAND experiment
 using reactor anti-neutrinos provide substantial support for the LMA region as shown in Figure~\ref{fig:KamlandExcPlot}. Future
 solar neutrino measurements
 will seek to improve on the definition of these neutrino properties as well as studying further the spectrum
 of electron
 neutrinos from the Sun, as discussed in more detail in the following section.

\begin{table*}
\caption{\label{tab:solardets} A summary of solar neutrino detectors
referred to in section II, restricted to experiments that have published results.}
\begin{ruledtabular}
\begin{tabular}{lllll}
Name & Dates & Location & Mass(Fid.) & Detector Details \\ \hline
Homestake~\cite{davis1} & 1968 - 2001 & Homestake mine, U.S.A. & 0.7 kton & Perchlorethylene \\
SAGE~\cite{SAGE} & 1990 - & Baksan Laboratory, Russia & 60 tons Ga & Gallium metal \\
GALLEX - GNO~\cite{GNO} & 1990 - & Gran Sasso Laboratory, Italy & 30 tons Ga & Gallium Chloride \\
Kamiokande III~\cite{KAMIOKA} &  1983-1996 & Kamioka mine, Japan &  4.5(0.68) kt $H_2$O & 1000 50-cm PMTs \\
Super-Kamiokande~\cite{SK} &  1996- & Kamioka mine, Japan & 50(22.5) kt $H_2$O & 11100 50-cm PMTs\\
SNO~\cite{SNONIM} & 1998- & Creighton mine, Canada & 1.1 (0.8) kt $D_2$O &  9456 20-cm PMT's\\
\end{tabular}
\end{ruledtabular}
\end{table*}

\section{Future solar neutrino experiments}

With the observation of flavor conversion by the combined analysis of 
SNO (CC) and Super-Kamiokande (ES), and by SNO(NC+CC) on its own, 
the scientific goals of upcoming experiments and future
projects have shifted from discovery and identification of the
solar neutrino flux deficit to a comprehensive study of the 
phenomena. The central challenge now is to measure, with high accuracy,
neutrino oscillation parameters and to determine accurately the neutrino 
fluxes of the primary solar pp-, pep- and $^7$Be branches.
Beyond this clear cut goal, and despite the apparent
consistency of observations with solar model predictions and 
oscillation scenarios, unexpected results might await and 
reveal further new physics. 

Details of stellar evolution theory can be tested by 
comparing experimental neutrino fluxes with 
theoretical predictions \cite{SSM1,SSM2}.  Most suited for 
such tests are the low energy pp-, pep- and $^7$Be fluxes, 
since they are predicted with rather small theoretical 
uncertainties compared with the high energy $^8$B: 
the pp- and pep-branch errors are estimated with 1\% and 1.5\% 
uncertainty, and the $^7$Be line with  10\% to be compared
with 18\% for the  $^8$B branch \cite{BP2000}.

\begin{figure}
\centering
\includegraphics[width=3.1 in]{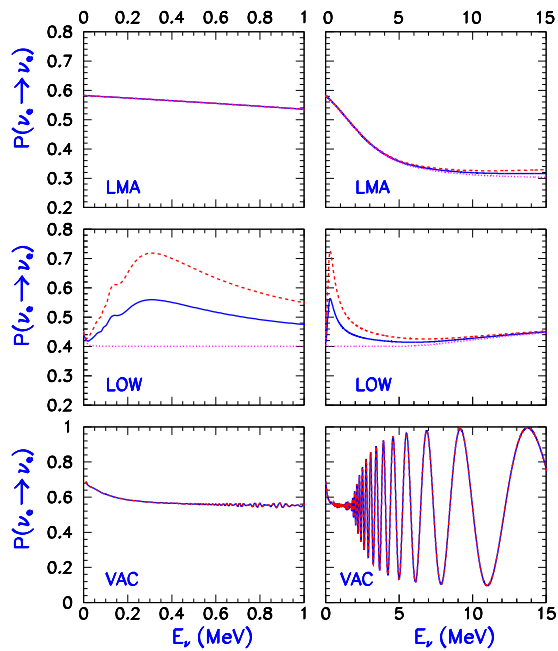}
        \caption{\label{lowe-survival}
	Energy dependent survival probabilities for solar 
	neutrinos for various oscillation scenarios. The 
	distinct differences at energies below 1 MeV should 
	be noted.
	The dashed, solid and dotted lines correspond to night, 
	average and day spectra respectively \cite{bgp2002}. 
Including the results from the KamLAND reactor neutrino experiment, all solutions but the LMA solution are excluded.}
\end{figure}

\begin{figure}
\includegraphics[width=3.6 in]{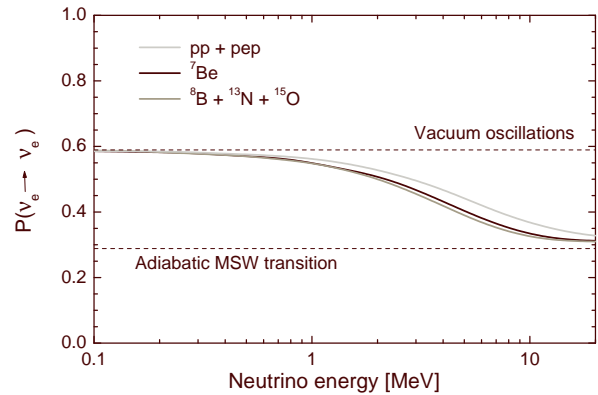}
\caption{\label{LMA-survival}
Electron survival probabilities for pp-, pep-, $^7$Be-, $^8$B-, $^{13}$N-
and $^{15}$O neutrinos for $\Delta {\rm m}^2 = 7.1 \cdot 10^{-5}
{\rm eV}^2 $ and $\sin^2 2\Theta = 0.41$.}
\end{figure}

Oscillation solutions derived from solar neutrino data alone,
allow two distinct parameter ranges, named LMA, LOW/VAC MSW-solutions and  
displayed in  Figure~\ref{fig:LisiMSW}. Including the recent results from the 
KamLAND experiment~\cite{KAMLAND} into the global analysis \cite{globalSol&KL} 
which reported evidence for the disappearance for  ${\bar \nu}_e$ from nuclear reactors, 
and provided that CPT symmetry is not violated 
(ie. no difference in the disappearance probability 
P(${\bar \nu}_e \rightarrow {\bar \nu}_x$) and  P(${\nu_e} \rightarrow {\nu_x}$),
all but the LMA solution are excluded (cf. Figure~\ref{fig:KamlandExcPlot}).
This solution makes distinct predictions for the electron neutrino
survival probabilities of approximately 60~\% for the
pp-, pep- and $^7$Be neutrino flux as diplayed in Figure~\ref{lowe-survival}.
Details of the  survival probabilities for
pp-, pep-, $^7$Be-, $^8$B-, $^{13}$N- and $^{15}$O neutrinos
as a function of energy are displayed in Figure~\ref{LMA-survival} 
\cite{hampel} for
$\Delta {\rm m}^2 = 7.1 \cdot 10^{-5} {\rm eV}^2 $ and
$\sin^2 2\Theta = 0.41$. The relative shift of the survival probabilities
for pp- and pep neutrinos with respect to eg. $^7$Be is related to
the different radial distribution of neutrino production inside of the
sun, and thus to the difference in electron density during propagation
of the neutrinos.


To address these fundamental issues of particle- and astrophysics, 
pp-, pep- and $^7$Be neutrinos which are emitted in the 
predominant terminations of the solar fusion process, 
will be studied with new detectors that provide information 
about the energy and time of the neutrino interaction.

\subsection{Real-time detection of low-energy solar neutrinos}

Rigorous tests of particle properties, and simultaneously of solar model 
predictions, require measurements of the neutrino spectra for each flavour separately.
The electron neutrino ($\nu_e$) flux
can be probed by the charged current inverse electron-capture 
reaction $\nu_e + (A,Z) \rightarrow e^- + (A,Z+1)^* $ (CC).
Solar neutrinos, converted to muon ($\nu_\mu$) and tau ($\nu_\tau$)
neutrinos can not produce their charged lepton partners, because
their restmasses are large compared to the energy available.
However, the $\nu_{\mu,\tau}$ flux component can be derived from 
the combined analysis of inverse electron-capture detection  with
elastic electron scattering  $\nu_{e,\mu,\tau} + e^-
\rightarrow \nu_{e,\mu,\tau}$ (ES), as shown by SNO and Super-Kamiokande.
The ES detection involves both charged and neutral
current interactions, albeit the contribution of the latter to the
cross section is only about 1/6. Therefore, high precision measurements
are needed, in particular if only a fraction of the flux
is converted to $\nu_{\mu,\tau}$'s.

The total flux of all active flavors - the quantity
to be compared with solar model predictions - can only be derived
by a combined analysis of CC and ES experiments.
An exclusive neutral current detection technique, as for 
the high energy $^8$B flux, is not at hand to be realized 
experimentally.

The Cherenkov technique for measuring solar neutrinos 
has been successfully used for $^8$B neutrinos.  However, at energies
at 1 MeV or below, the Cherenkov photon yield is 
insufficient to perform spectroscopic measurements. For this reason,
scintillation techniques will be used for the upcoming 
low-energy neutrino experiments. A primary
photon yield of up to $1\times 10^4$/MeV  can be achieved 
for organic liquid scintillators and up to  $4\times 10^4$/MeV 
in liquid noble gases. Ultra pure liquid scintillators
will be used for ES detection, while metal loaded organic
scintillators for CC detection. 
The use of a Time Projection Chamber (TPC) is under study as an alternative to scintillation detection
for ES with the goal to reconstruct the full kinematics of the ES reaction.

Only a few promising candidate nuclei exist for CC real time
detection.
Either the transition must populate an isomeric excited state
with a subsequent electromagnetic de-excitation,
or the final state needs to be unstable in order  to provide a 
coincidence tag to discriminate against background events
such as ES signals and radioactive decays.  
Moreover, the energy threshold of the transition needs to be 
sufficiently low to have sensitivity to sub-MeV neutrinos. 
Only a few isotope meet these requirements of which 
$^{115}$In, $^{100}$Mo, $^{82}$Se, $^{160}$Gd, $^{176}$Yb 
are under careful consideration.

The SSM interaction rates (without flavor conversion) 
for ES experiments amount to about 0.5/ton/day
for $^7$Be-$\nu$'s (860 keV branch), to 2/ton/day for pp-neutrinos 
and 0.04/ton/day 
for pep-neutrinos. Hence target masses of  about 100 ton
are required for $^7$Be- and pep-measurements, while about
10 tons are sufficient for pp-neutrinos in order to acquire
a rate of ten events per day.

The interaction rates for CC detection via inverse electron capture 
depend on the nuclear structure of the 
isotope under consideration and its relative abundance. For example,
the interaction rate in  indium ($^{115}$In, 
matrix element $B(GT)$=0.17, 95.7\% natural abundance) 
is 0.07/ton/day for $^7$Be-$\nu$'s and 0.3/ton/day for 
pp-$\nu$'s assuming SSM fluxes without oscillations.

\subsection{Backgrounds to neutrino detection at low-energies}

The main challenge of real time neutrino detection at low 
energies are backgrounds from radioactive decays. 
Primordial radioactive contaminants present in detector materials 
such as  $^{40}$K, $^{238}$U, $^{232}$Th and 
their progenies, as well as anthropogenic $^{85}$Kr,
cosmogenic $^{39}$Ar, and radiogenic $^{14}$C 
dominate the detector signal if not removed 
carefully. 

Scintillation based ES experiments rely on the detection of a single recoil 
electron. Therefore, any radioactive decay with similar energy
deposition can mimick a neutrino event. Concentration of radioactive
elements need thus to be $< 1 \mu \rm{Bq/ton}$ to allow a 
signal/background ratio $>1$. 
This translates to contamination limits 
$\lsim 10^{-16}$g~U(Th)/g for  $^7$Be detection.
CC experiments typically require less strigent limits
since delayed coincidence tags suppress backgrounds.

Further backgrounds arise from cosmic ray muon induced
radio isotopes. To minimize this interference, detectors are 
located deep underground.
A rock overburden, for example, of 3400 mwe (meters of water equivalent), as encountered at the Gran Sasso
underground laboratories,
reduces the muon flux to 1.1~h$^{-1}$~m$^{-2}$. 
Despite a reduction of a factor $10^6$ with respect to the sea level 
flux, the in-situ production of radio-isotopes 
by spallation reactions is still of relevance. For example, 
backgrounds to pep-neutrino detection via ES at this depth 
is dominated by in-situ production of $^{11}$C (t$_{1/2}=20.4$~min)  
by muons and their secondary particles. 
The $^{11}$C production (and decay) rate amounts
0.15/ton/day to be compared to 0.04/ton/day pep-neutrino 
interactions \cite{NA54}. Therefore, efficient spallation cuts, or alternatively, 
a deeper underground location are required to extract a pep-neutrino signal.

\subsection{Upcoming experiments}

The Borexino experiment is the pioneering project for real time
solar neutrino spectroscopy at low energies  \cite{BXAP16}. 
The construction of the experimental
installations is nearing completion in the underground 
laboratories at Gran Sasso, Italy (LNGS). Startup of the 
experiment is expected for 2004. 
The primary goal of Borexino is to measure the 0.86~MeV $^7$Be-$\nu$
line via elastic neutrino--electron scattering (ES).
Further physics goals include the detection of pep-neutrino, the 
low energy part of the $^8$B spectrum, neutrinos from supernovae,
as well as anti-neutrinos (${\bar \nu}_e$) from 
distant nuclear reactors and from geophysical sources.

\begin{figure}
\centering
\includegraphics[width=9cm]{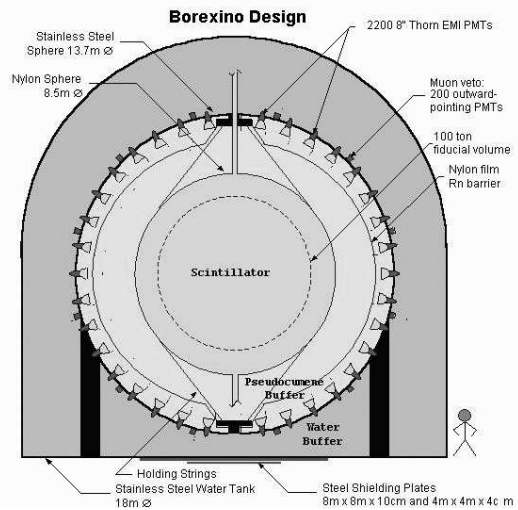}
        \caption{\label{bxdet} 
	Schematic view of the Borexino detector at Gran Sasso.
	A stainless steel sphere confines 
	the inner detector (ID) containing about 300 ton liquid scintillator
	and 1 kton of transparent buffer liquid.
	Scintillation photons are detected by 2200 photomultiplier 
	giving energy and location of the interaction. 
	A fiducial central mass of 100 ton can 
	be determined in the off-line analysis.
	The steel sphere is housed in a water tank equipped with 
	210 photomultiplier. The outer detector (OD) 
	serves as a shield against ambient radiation as well 
	as a muon track detector.}
\end{figure}

Fig.~\ref{bxdet} displays schematically the 
Borexino detector. Neutrino detection occurs via ES in an 
ultra-pure liquid scintillator target confined in a transparent
nylon vessel. The scintillator, with a mass of 300 tons (100 tons fiducial volume) is composed of 
pseudocumene (PC) and PPO at a concentration of 1.5~g/l. 
About $10^4$ primary photons/MeV
are emitted with a wavelength distribution peaked at 380 nm.
Photon detection is realized with 2200 photomultipliers 
(8'',  30\% coverage) at single photo-electron threshold
providing a  yield of $\gsim 400$~photo-electrons/MeV.
An energy resolution of  $\gsim 5$\% ($1\sigma $) at 1~MeV is therefore
attainable. The location of the interaction within the detector 
is determined with an uncertainty of $\lsim 10$~cm ($1\sigma $) 
at 1~MeV using the time-of-flight method of the photons.

A trigger is generated by $\approx 15-20$ PMT hits occurring in a time
window of 60 ns corresponding to a threshold of
about 50~keV energy deposition.
The analysis threshold for recoil electrons from  $^7$Be-$\nu$
ES will be at 250~keV, depending on the actual $^{14}$C activity
in the scintillator.

To achieve the ultra-low background rate within the fiducial
inner volume, the detector has an onion-like structure 
with increasing radio purities from outside to inside.
$\gamma$-rays coming from detector components outside the 
scintillation volume are  attenuated by the buffer liquid 
surrounding the active volume. The final reduction of 
external activity is achieved by the outer layer of the 
liquid scintillator (``self-shielding''), 
defining a fiducial mass of 100 tons.

The cumulative background rate internal to the liquid scintillator 
must be $\lsim 1\times 10^{-6}\,{\rm s}^{-1}\, {\rm m}^{-3}$ in the
neutrino analysis window between 250~keV and 800~keV.
This translates to limits 
$\lsim 10^{-16}$g/g
for uranium and thorium,
and their progenies (assuming secular equilibrium), 
to $\lsim 10^{-14}$g/g for potassium,
to $\lsim 10^{-10}$g~/g for argon  
and to $\lsim 4\times 10^{-16}$g/g for krypton.
Due to the high mobility of radioactive noble gases, in particular 
$^{222}$Rn, $^{39}$Ar, $^{85}$Kr, ultra-high vacuum leak-tightness 
standards of the system is required and gases in contact with the liquid 
scintillator, such as nitrogen need to be purified.
A further source of backgrounds comes from surface deposition of 
$^{222}$Rn progenies on detector components which 
are subsequently in contact with the liquid scintillator. 
In particular the buildup of 
$^{210}$Pb (t$_{1/2}=22.3$~yr) must be controlled
since it feeds the decays of $^{210}$Bi (t$_{1/2}=5$~d)
and $^{210}$Po (t$_{1/2}=138$~d) which create signals 
in the neutrino energy window.

All materials have been screened with high-purity germanium spectroscopy,
and selected for low radioactive trace contaminants \cite{BXAP18}.
Surfaces which are in contact with the liquid scintillator
have been tested for radon emanation, and are specially 
treated to remove $^{210}$Pb and $^{210}$Po deposits.

The detector concept and in particular, the attainable  
radioactive trace contaminations have been studied in a 
pilot experiment, the Counting Test Facility at Gran Sasso (CTF) \cite{ctf}.
During construction and startup of Borexino, the CTF serves 
as a sensitive instrument to verify the performance of the ancillary 
plants, such as the liquid handling distribution, the various 
purification systems, as well as the purity levels of PC 
prior to filling the Borexino detector.

A similar detector, however larger in size, 
has been realized by the KamLAND collaboration
in the Kamioka mine in Japan \cite{KL}. The main objective of this
experiment is to probe the oscillation parameter space of the
solar MSW large mixing  solution with electron 
anti-neutrinos (${\bar \nu}_e$) from distant nuclear reactors.
Radio purity requirements for ${\bar \nu}_e$-detection are 
less stringent compared to solar neutrino detection via ES.
Data taking with reactor neutrinos started early 2002 and first
results have been published~\cite{KAMLAND}.
In a second phase, with an upgraded liquid handling and purification system, 
it is intended to measure solar $^7$Be
neutrinos, provided that trace contaminations
are at levels as required for Borexino. 

The active volume of the KamLAND detector consists 
of 1000 ton of liquid scintillator composed of 
PC, mineral oil (dodecane) and PPO (1.5 g/l). 
An energy resolution of $\sim7$\% ($1\sigma$) at 1 MeV 
has been reported and is expected to improve to $\sim6$\%
in the final detector configuration.
First results on $^{238}U$, $^{232}$Th were $(3.5\pm 0.5) \times 10^{-18}$ g/g,
$(5.2 \pm 0.8) \times 10^{-17}$ g/g, while impurities such as 
$^{85}$Kr and $^{210}$Pb still need to be reduced substantially
to allow $^7$Be-$\nu$ detection.

\subsection{Next generation experiments}

\begin{table*}
\caption{\label{tab:projects}
	R\&D projects for sub-MeV solar neutrino detection.
	ES: Elastic scattering of neutrinos off 
	electrons (CC+NC), 
	CC: neutrino capture (charged current; CC); 
	LS: liquid scintillator; CH: hydro-carbon}
\begin{ruledtabular}
\begin{tabular}{lll}
	\hline Project & Method & Technique (Target) \\
	\hline LENS~\cite{lens} & CC: $^{115}$In,$^{176}$Lu & LS (CH+metal) \\
	MOON~\cite{moon} & CC: $^{100}$Mo & hybrid or LS (CH+metal)\\
	XMASS~\cite{xmass} & ES  & LS (Xe) \\
	HERON~\cite{heron} & ES  & LS (He)   \\ 
	CLEAN~\cite{clean} & ES  & LS (He,Ne)\\
	TPC~\cite{tpc} & ES  & TPC(He,CH) \\
	\hline
\end{tabular}
\end{ruledtabular}
\end{table*}

$^{14}$C/$^{12}$C ratios at 10$^{-18}$
in organic scintillators, as determined with the CTF \cite{C14},
prohibit the measurement of pp-neutrinos in ES experiments 
since the 156 keV $\beta$-decay endpoint significantly 
obscures the pp-neutrino energy range. A promising approach to overcome
this background is to avoid organic liquids and instead, to use 
liquefied noble gas as a scintillator. Projects with helium, neon
and xenon are under investigation and are listed in 
Table~\ref{tab:projects}.
Recent development for pp- and $^7$Be CC detection  
focuses on the isotopes  $^{115}$In (LENS) and on $^{100}$Mo (MOON).  
Most advanced amongst the various R\&D projects, listed in 
Table~\ref{tab:projects}, are XMASS for ES-
and LENS for CC detection.
In the following, experimental and conceptional 
details will be outlined for these two projects, 
exemplary for  the next generation 
of experiments. Further details of recent progress 
of the various projects can be found at \cite{lownu2002,taup01}.

The XMASS \cite{xmass} collaboration pursues the concept of  
pp- and$^7$Be-$\nu$ detection via ES in a liquid xenon
scintillation detector. High density liquid xenon (3.06~g/cm$^3$) 
provides efficient self shielding and a compact detector design. 
A geometry similar to that
of Borexino, but smaller in size, could thus be realized. The scintillation 
photons (175 nm) will be detected with
newly developed low-background photomultiplier tubes (Hamamatsu)
consisting  of a steel housing and a quartz window
at liquid xenon temperatures ($\lsim 165$~K). 
Main sources of backgrounds arise from $^{85}$Kr $\beta$-decay and 
from $2\nu-\beta\beta$ decay of $^{136}$Xe. The first isotope must be 
less than $4\times 10^{-15}$gKr/g. If the $\tau_{1/2}$ of $^{136}$Xe
is $\lsim 8\times 10^{23}$~years, as theoretically expected, 
then isotope separation is needed.
A prototype detector containing 100 kg of liquid Xe is under
construction. Milestones during this R\&D phase
will include the determination of the $2\nu-\beta\beta$ half-life
and the optical properties of liquid xenon, such as the 
scattering length of scintillation light. 

Various candidate nuclei are being investigated by the LENS \cite{lens} collaboration.
The ongoing research focuses now on  $^{115}$In loaded into an 
organic liquid scintillator with 5-10\% in weight.    
Complexing ligands under study include carboxylic acids, 
phosphor organic compounds as well as chelating agents.
A detector containing about 10 tons of indium is under consideration. 
In order to discriminate against backgrounds, dominated 
by the $\beta$-decay of $^{115}$In 
($Q_\beta = 496$~keV, 0.26 Bq/g of In) and the accompanying Bremsstrahlung,
a detector with high spatial granularity is required. This
will be realized by a modular design. A basic module
has a parallelopiped shape with a length of 2-3 m, determined by
the absorption length of the liquid scintillator.
The cross section of the module varies from $5\times 5$~cm$^2$
to $10\times 10$~cm$^2$ depending on metal loading, scintillator 
performance and further optimization criteria under study.

The nuclear matrix element $B(GT)=0.17$, relevant for $\nu_e$ interaction, 
has been determined via the (p,n) reaction. This value is 
sufficiently accurate to evaluate the target mass necessary 
for the LENS detector. However, to derive the neutrino fluxes
with accuracy $\sim 5$\%, it is planned to use an 
artificial $^{51}$Cr neutrino source of several MCi strength 
to determine the neutrino capture cross section.

In order to study the detector performance as close as possible to 
the final detector geometry, the LENS Low-Background-Facility (LLBF)
has been newly installed underground at the Gran Sasso Laboratories (LNGS). A low-background passive 
shielding system with 80 tons of mass, located in a clean room, 
can house detector modules with dimensions up to 70~cm $\times$ 70~cm 
$\times$ 400~cm. All shielding materials have been selected in order 
to minimize the intrinsic radioactive contamination. First results 
from the prototype phase are expected end of 2003.

\subsection{Outlook}
In near future, the detection of $^7$Be and, conceivably pep-neutrinos
via ES will be addressed by Borexino and by 
KamLAND.  The next generation projects still have to pass the
threshold from promising ideas to feasible experiments.
Results from the ongoing prototype activities will 
show within the very near future which of the projects 
have the potential to be realized. On the long term, 
complementary measurements via ES- and 
CC detection are desirable for all neutrino branches in order
to scrutinize solar models as well as neutrino properties.


\section{Supernova Neutrinos}

Up to 99~\% of the roughly $10^{53}$ ergs of energy released in type II and type
Ib supernova is carried away by neutrinos. Many of the currently operating
neutrino telescopes are capable of detecting these neutrinos and providing
detailed information on such supernovae. The standard model for a type II
supernova is a massive star that has reached a point in its evolution when
the pressure from fusion reactions can not support the gravitational pressure
of the outer regions. An inward collapse is initiated and proceeds until the
central density reaches nuclear density at which time a "bounce" occurs and an
outward shock wave is created. The interaction of this outward moving shock
wave with the outer material contributes to an explosion, aided by energy
deposited behind the shock wave from neutrinos created in the collapsing
core. 

\begin{figure}
\includegraphics[width=3.5in]{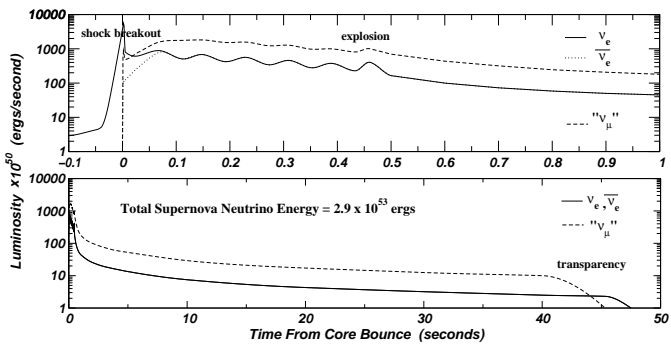}
\caption{\label{fig:BurrowsLuminosity}
(a) The luminosity (in ergs per s) vs time (in ms) for the $\nu_e$, ${\bar \nu}_e$ and collectively,
the $\nu_\mu$(${\bar \nu}_\mu$) and $\nu_\tau$(${\bar \nu}_\tau$) during the first second of the neutrino burst
in the generic model employed by Burrows et al \cite{Burrows}. (b) Same as (a) but for the first 50 s. }
\end{figure}

Although there are many detailed models of supernova dynamics, the models
share a number of generic features. These are illustrated in Figure~\ref{fig:BurrowsLuminosity}
from an article by Burrows et al~\cite{Burrows} that explores the potential for
supernova neutrino detection by a number of different detectors. Neutrinos
and anti-neutrinos of all active types are produced during a period lasting
up to about 10 seconds. The average neutrino energies are typically 12 MeV
for electron neutrinos, 15 MeV for electron anti-neutrinos and 18 MeV for
the muon and tau neutrinos and anti-neutrinos (labelled as mu neutrinos in the figure).
The time period for neutrino
generation can be divided into three principal regions, a collapse phase,
an accretion phase and a cooling phase following an explosion. In the
collapse phase electron capture on protons produces a sharp burst of
electron neutrinos. When the core bounce occurs,
neutrinos and antineutrinos are produced in the
high temperature wake of the launched shock wave. Throughout the accretion
phase, there is gradual spectral hardening of all the neutrino species that
is increased sharply as an explosion occurs, reversing the flow of in-falling
matter and beginning the cooling phase during which the spectra soften again.

\begin{table*}
\caption{\label{tab:SNDetection}
Signal totals for various processes in representative supernova detectors at 10 Kpc~\cite{Burrows}. $\nu_\mu$ represents
 $\nu_\mu$, $\nu_\tau$, ${\bar \nu}_\mu$, ${\bar \nu}_\tau$.}
\begin{ruledtabular}
\begin{tabular}{llll}
Reaction & SNO & Superkamiokande & LVD\\
\hline 
${\bar \nu}_e$ p &  331& 5310 & 342\\
$\nu_e$$e^{-}$ & 12 & 77.6  & 8.04\\
${\bar \nu}_e e^{-}$& 6 & 16.9 & 1.49\\
$\nu_\mu e^{-}$& 7 & 49.9 & 3.28\\
$\nu_e$O (CC)& 3 & 36.0 & - \\
${\bar \nu}_e$O (CC) & 3 & 45.5 & - \\
$\nu_x$C & - & - & 22\\
$\nu_e$d (CC) & 81.9 & - & \\
${\bar \nu}_e$d (CC) & 66.7 & - & \\
$\nu_e$d (NC) & 35.2 & - & \\
${\bar \nu}_e$d (NC) & 37.2 & - & \\
$\nu_\mu$d (NC) & 200 & - & \\
\end{tabular}
\end{ruledtabular}
\end{table*}

\begin{figure}
\includegraphics[width=3.5in]{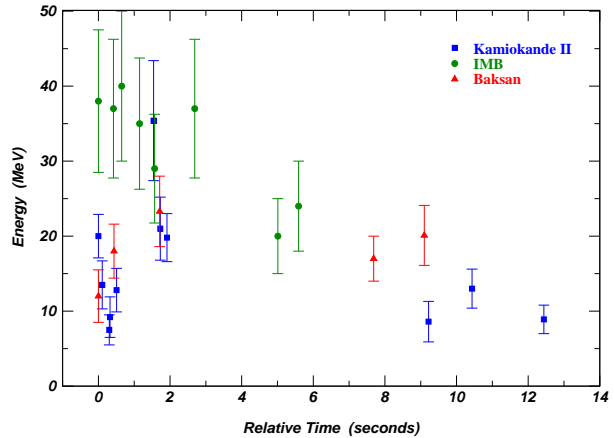}
\caption{\label{fig:SN1987a}
Electron/positron energies and relative times for the events identified by the Kamiokande II, IMB,
and Baksan detectors during SN1987a. The time of events recorded by each detector is relative to
 the first event in that detector~\cite{Jaret}.}
\end{figure}

There are many detectors that have the capability to observe neutrinos
from a Supernova in our galaxy or the Large Magellenic Cloud (LMC). This
was demonstrated very dramatically when supernova 1987a was observed in
three underground detectors. The Kamiokande, IMB and Baksan detectors
observed neutrinos from this event that occurred in the LMC, about 51
megaparsecs from Earth~\cite{SN87A-Gen,SN87A-Kam,SN87A-IMB,SN87A-Bak,SN87A-phys}.
Figure~\ref{fig:SN1987a} shows the numbers, energies and arrival
times of these neutrino events~\cite{Jaret}.

The detection process for these detectors is dominated by anti-electron
neutrinos interacting with protons in water (2350 tons fiducial volume
in Kamiokande, 5500 tons in IMB) or liquid scintillator (200 tons in
Baksan). (A further 5 events were observed by the LSD detector but they
occurred about 5 hours out of time with the other detectors.) The 
interactions on protons in these detectors has about ten times higher
probability than elastic scattering from electrons, leading to the
assignment of these events to this process with the highest probability.
The events have been carefully analyzed and compared with models for
supernova neutrino emission, with good agreement in general \cite{SN87A-phys}.

Experiments presently in operation or planned for the future have a broader
capability for the detection of the next nearby supernova. The
Super-Kamiokande detector has a fiducial volume of 32 ktons of light
water and would provide significant signals for anti-electron neutrinos
interactions with protons and all neutrino flavors interacting by elastic
scattering on electrons. The SNO detector has a fiducial volume of 1.1 kton
of heavy water and 1.5 kton of light water and can detect all neutrino
types via several reactions. The LVD detector is a liquid scintilator detector with 1~kton 
fiducial mass. Supernova neutrinos are detected mainly through inverse beta decay:
$\overline{\nu}_e p \rightarrow e^+ n$. In addition, neutrinos of all flavor
can be detected by neutral current interactions off carbon, which produce
mono-energetic 15.11~MeV gamma rays.  Several other scintillator based detectors are in operation that are
sensitive to inverse beta decay, including miniBooNE \cite{mboone} and KamLAND. 

The combination of detectors
listed in Table~\ref{tab:SNDetection} from Burrows et al.~\cite{Burrows} can provide
a detailed measure of the flavor composition
of the neutrino flux as a function of time, and energy to test models of
supernovae. Neutrino flavor change processes can also occur during supernova
and can be studied with the array of detectors now in operation.
Because neutrinos can precede light emission from supernovae by as much as a
few hours, it can be valuable to provide an advance signal from neutrino
telescopes to other astronomical telescopes. Several of the existing
experiments have formed a coordination group~\cite{SNEWS} known as the SuperNova Early
Warning System (SNEWS) to look for coincident bursts of events among several
of the experiments and provide as rapid a warning to other telescopes as
possible. Whereas the timing of the leading edge of the neutrino burst
provides only modest localization through triangulation~\cite{angdist}, several of the
experiments (SuperKamiokande and SNO) could provide localization to within
a few degrees for a supernova near the galactic centre via the highly
directional neutrino-electron elastic scattering reaction.

Few supernovae have been observed visually in our Galaxy in the past-
about 6 in the past 1000 years. This is thought to be because most
of them are obscured by intergalactic dust. Predictions of supernovae
rates in our Galaxy range from one per 10 years to one per 100 years.
The techniques used to make these estimates include estimates of stellar
evolution rates, estimates based on iron abundances produced in supernovae,
studies of supernovae rates in other galaxies and studies of supernova
remnants. Detectors currently in operation have volumes providing sensitivity
only out to the LMC region. A widely accepted range for the occurence rate is about one per 30 to 50
years in our galaxy and the LMC combined~\cite{BH-cutoff, Nature}. Detectors capable of reaching far
enough to obtain a rate of about one supernova per year have been discussed,
but a volume more than 100 times the present detectors would have to be
instrumented with a detection threshold on the order of about 10 MeV.
Arrays of neutron detectors have also been considered but the total volume
is a daunting task. 

Large neutrino detectors may also be sensitive to so-called relic supernova
neutrinos. These originate from sum of all distant supernovae, and appear as
a relatively low background of interactions rather than as the brief burst of
events characteristic of a nearby core collapse. Detailed predictions depend
on the cosmological model that determines the average redshift and epoch of
maximum supernova rate~\cite{relicmodels}, and include the effects of neutrino
oscillation. As with nearby supernovae, the most promising reaction is through
the inverse beta decay reaction of ${\bar \nu}_e$. The flux of ${\bar\nu}_e$
is predicted to be rather small: less than 50 $\rm cm^{-2} s^{-1}$ over the
entire spectrum. This is $10^5$ times smaller than the ${\rm ^8B}$ solar
neutrino flux, which therefore offers a sizeable background.

In searches performed so far, the signal is sought in an energy window above
the ${\rm ^8B}$ endpoint of 18 MeV and below the onset of numerous
atmospheric neutrino interactions around 50 MeV. In this window there are
still challenging backgrounds due to Michel electrons from untagged muon
decays (such as from $\nu_\mu$ interactions with the muon below a Cherenkov
detection threshold), and low energy atmospheric $\nu_e$ interactions. The
most recent studies are from Super-Kamiokande~\cite{skrelic}, which performed a combined
signal and background fit to the energy spectrum in this region.  No
appreciable signal was detected in 92 kton-years and they limit the flux of
relic supernova ${\bar \nu}_e$ to be less than 1.2 $\rm cm^{-2} s^{-1}$ above
$E_\nu > 19.3$ MeV.

\begin{table*}
\caption{\label{tab:dets} A summary of atmospheric neutrino detectors, restricted to experiments that 
have published results on a sample of at least 100
atmospheric neutrinos. If a detector was upgraded, the final
name and configuration is listed; however the earliest start date for
any version of the experiment is provided. The first section lists fine grained
tracking detectors and the second section lists water Cherenkov detectors.}
\begin{ruledtabular}
\begin{tabular}{lllll}
Name & Dates & Location & Mass(Fid.) & Detector Details \\ \hline
Baksan~\cite{baksan} &  1978- &  Baksan Laboratory, Russia&  0.33(-) kt & Liq. scintillator tanks \\
Frejus~\cite{frejus} &  1984-1988 &  Frejus tunnel, France/Italy & 0.9(0.7) kt &  \parbox{1.5in}{Flash/geiger + iron planes} \\
Soudan 2~\cite{soudan} &  1989-2001 & Soudan iron mine, U.S.A. & 0.96(0.77) kt &  Drift tubes + corrugated steel \\
MACRO~\cite{macro} &  1988-2001 & Gran Sasso Laboratory, Italy &  4.7(-) kt &  Streamer tube + liq. scint. \\ \hline
Kamiokande III~\cite{kam} &  1983-1996 & Kamioka mine, Japan &  4.5(1.0) kt & 1000 50-cm PMTs \\
IMB 3~\cite{imb} &  1982-1991 &  Morton Salt mine, U.S.A. &  8.0(3.3) kt &  2048 20-cm PMTs\\
Super-Kamiokande~\cite{sk} &  1996- & Kamioka mine, Japan & 50(22.5) kt &  11100 50-cm PMTs\\

\end{tabular}
\end{ruledtabular}
\end{table*}

\section{\label{sec:atmnu}Atmospheric Neutrino Detectors}

In the early 1980's, the first massive (of order 1~kton) underground detectors
were constructed, primarily to detect proton decay with a lifetime of
less than 10$^{32}$~years, as predicted by early Grand Unified
Theories~\cite{pati,georgi}. The most serious background for proton decay
searches is from atmospheric neutrinos, at a rate of approximately 100
events/yr/kt. Atmospheric neutrinos are produced as decay products in
hadronic showers resulting from the collision of cosmic rays with nuclei in
the upper atmosphere. The study of atmospheric neutrinos is interesting apart
from background studies, as they provide a sample of muon and electron
neutrinos that travel distances ranging from 10 km to 13000 km, thereby
providing for the study of neutrino oscillation. Table~\ref{tab:dets}
summarizes some of the experiments to date that have studied atmospheric
neutrinos.

Atmospheric neutrinos interact with the nucleus, at a typical energy of 1
GeV, although the detectable spectrum extends beyond 1 TeV. In order to
reject background from cosmic ray muons, as well as to reconstruct
the details of the event cleanly, the vertex position of the neutrino interaction is
typically required to be within some fiducial volume. There are several
standard categories of events determined by the extent of the final state
particles.  Low energy events, around 1 GeV have all of the final state
particles ``fully contained'' in the detector.  Electromagnetic showers, even
at energies of 10-100~GeV, are also usually fully contained.  By placing
energy or momentum cuts on the events, one may statistically define
sub-samples of lower or higher average neutrino energy. A cut around 1 GeV
(suitable for identifying proton decay), has traditionally demarked
``sub-GeV'' and ``multi-GeV'' data samples.  Higher energy charged current
$\nu_\mu$ interactions may result in the muon exiting the detector; these are
referred to as ``partially contained''. Some of the experiments are equipped
with outer detectors (also referred to as veto- or anti-detectors) to easily
identify exiting particles, as well as to reject cosmic ray background.

The other possible background source for contained events are high
energy neutrons from muon interactions nearby the detector; this is a
potentially more serious background, because neutrons enter a detector
without any visible tracks. It has been reported that such neutron
backgrounds may be controlled by a sufficiently thick active shield;
 one may then veto nearby cosmic ray muons that produce neutrons~\cite{kam-pi0,s2-neutron}.

There is a third category of charged current $\nu_\mu$ events, where the
interaction occurs outside the detector, and the muon enters and either
passes through the detector or stops in the detector. These are referred to
as ``upward-going muons'' because one generally requires they originate from
below the horizon to ensure that a sufficient amount of rock absorbs ordinary
cosmic ray muons (at very deep locations, especially with a flat overburden,
the zenith angle requirement can be somewhat above the
horizon~\cite{sno-upmu}). Upward-going muons represent the highest energy
neutrinos ($\sim 1$ TeV) detected by atmospheric neutrino experiments, and
have good pointing resolution ($\sim$ few degrees) with respect to the
original neutrino direction. Therefore this sample is used for traditional
neutrino telescope studies such as the search for astrophysical point sources~\cite{astronu}
 of high energy neutrinos including short duration sources such
as gamma ray bursts~\cite{astrogrb}. In addition, dark matter annihilation in
the sun, earth, and galactic center may produce neutrinos and the signal has
been sought using upward-going muons~\cite{astrowimp}. All results to date
have been negative, i.e. statistically consistent with only the expected
background from atmospheric neutrinos (see also Section VI).

\begin{figure}
\includegraphics[width=3.3in]{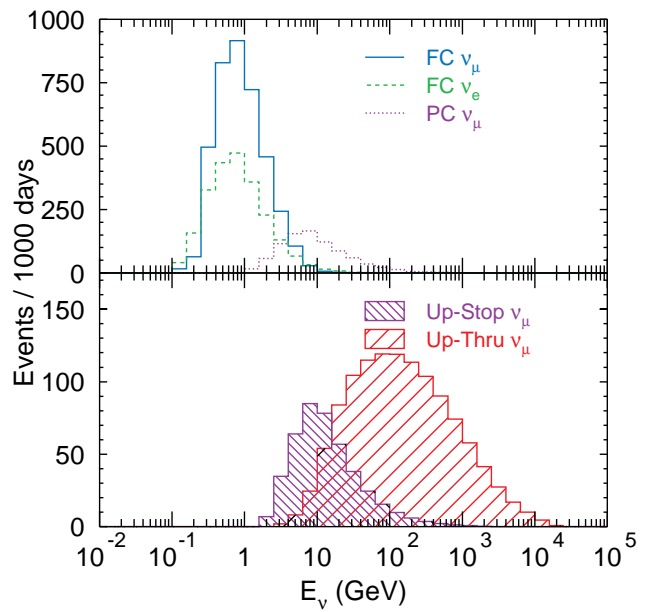}
\caption{\label{fig:spectra}
  The parent neutrino energy for several different classifications of
atmospheric events. These distributions are for the Super-Kamiokande
analysis; for other detectors they will be different depending on
detector size, which controls the maximum energy of a fully-contained event
and the minimum energy of a through-going muon.}
\end{figure}

Figure~\ref{fig:spectra} shows the distribution of parent neutrino energies
for fully contained, partially contained, and upward-going muon event
samples, specifically for the Super-Kamiokande analysis.

\subsection{\label{sec:watchen}Water Cherenkov Detectors}

\begin{figure}
\includegraphics[height=3.4in,angle=270]{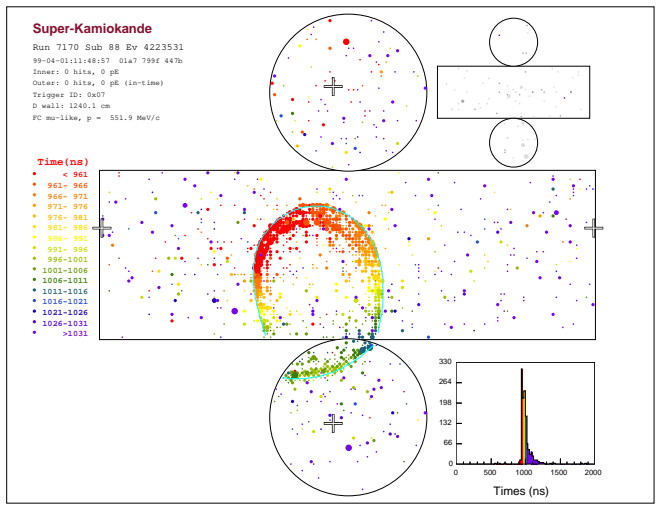}
\caption{\label{fig:sk-event}
An example event display from the Super-Kamiokande detector, showing
the sharp-edged Cherenkov ring from a 0.6 GeV muon. Each small circle
represents one PMT over threshold, with the size of the circle
correlated with the number of photoelectrons, and the color of the
circle related to the arrival time of the Cherenkov light.}
\end{figure}

Atmospheric neutrino events are detected in water Cherenkov detectors
by observing Cherenkov radiation from relativistic charged particles
in the final state of the neutrino interaction. A two dimensional
array of photomultiplier tubes on the
inside surface of the detector detects the photons. The hit time and
the pulse height from each PMT are recorded. The hit time,
with a typical resolution of a few ns for a single photo-electron
pulse, is used to reconstruct the vertex position.  The total
number of photo-electrons gives information on the energy of the
particles above Cherenkov threshold. It is conventional to refer to
``visible energy'', which is defined as the equivalent energy assuming
all of the Cherenkov light arises from electromagnetic showers.  Thus,
a muon of 1 GeV would produce about 0.9 GeV of visible energy, but an
electron of 1 GeV would produce 1 GeV of visible energy.
Figure~\ref{fig:sk-event} shows a fully-contained atmospheric neutrino
event observed in the Super-Kamiokande detector. 

It is possible to use the pattern of Cherenkov light to identify separately the
electrons and muons produced by charged current $\nu_e$ and $\nu_{\mu}$
interactions respectively. An electron produces an electromagnetic shower
while propagating in a medium, and the low-energy electrons and positrons
undergo multiple scattering while producing Cherenkov light. On the other
hand, a muon propagates in a nearly straight line, losing energy by the
ionization. In addition, the opening angle of the Cherenkov radiation is
smaller for a low energy muon. These differences are used to form a
likelihood for the Cherenkov pattern for electrons and muons. The
nomenclature used is either $e$-like and $\mu$-like or showering and
non-showering.

The first generation water Cherenkov detectors were IMB and
Kamiokande. IMB (originally named for Irvine-Michigan-Brookhaven)
started taking data in 1982 and ended in 1991 after two major
upgrades. The first phase of the detector (IMB 1) was equipped with
2048 12.5-cm diameter PMTs. The light collection was improved by the
addition of 60~cm $\times$ 60~cm $\times$ $1.3$~cm waveshifting
plates~\cite{imbwls}, for a brief running period (IMB 2).  Finally, the
detector was equipped with 2048 20-cm PMTs, again with a wave length
shifter plate (IMB 3).  The Kamiokande detector (the name is based on
Kamioka Nucleon Decay Experiment) used 1000 PMTs with 50-cm
diameter, a photocathode coverage of 20\%. For the second phase,
Kamiokande had an outer detector with thickness of the water between
0.5 and 1.5~m. The outer detector made it possible to identify
partially contained events. These experiments showed that the
observed fraction of muons relative to electrons was much smaller than
the Monte Carlo prediction~\cite{kamR,IMBR}. 
The Kamiokande result for multi-GeV and
partially contained events suggested a zenith angle dependence
indicative of neutrino oscillation \cite{kam-mgevpc}.

\begin{figure}
\includegraphics[width=3.1in]{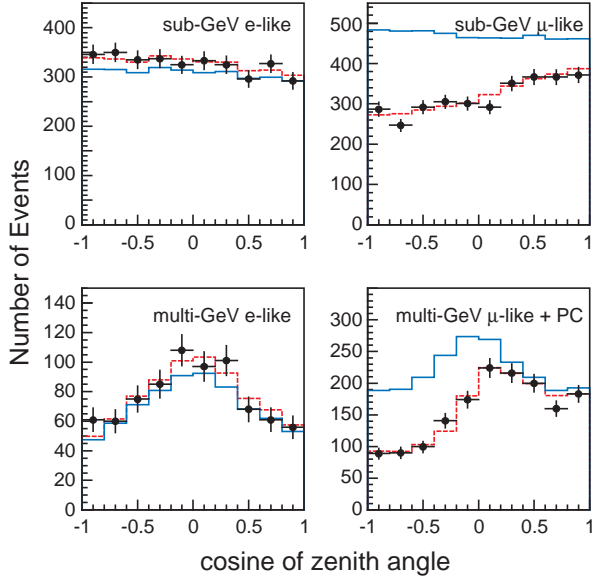}
\caption{\label{fig:zenith}
Zenith angle distributions observed in Super-Kamiokande
for sub-GeV $e$-like events(top-left),
for multi-GeV $e$-like events(bottom-left),
for sub-GeV $\mu$-like events(top-right), and
for multi-GeV $\mu$-like plus partially-contained events(bottom-right). The solid
lines show simulations for no neutrino oscillations, the dashed lines show the best fit
assuming oscillations.}
\end{figure}

The current generation water Cherenkov detector is Super-Kamiokande (see
Fig.~\ref{fig:skdet}), whose results dominate our understanding of
atmospheric neutrinos. It began taking data in 1996. Super-Kamiokande has a total mass
of 50~ktons, with an inner detector 36~m in height and 34~m in diameter. It
uses 11146 50-cm diameter PMTs, with photocathode coverage of 40\% of the
inner detector surface, a factor two higher than that of Kamiokande. An outer
detector surrounds the inner detector with 2~m thickness of water, equipped
with 1885 20-cm diameter PMTs with wavelength shifting plates (these were recovered
from the IMB experiment).  To limit the scattering of the Cherenkov photons,
Super-Kamiokande has an extensive water purification system. As a result, the
attenuation length of the water is longer than 100~m for 400~nm wavelength light.
The fiducial volume for neutrino vertices is 2 m from the plane of
photomultiplier tubes, resulting in a 22.5~ktons mass. The large mass and
photocathode coverage allows for high statistics and detailed studies of
atmospheric neutrinos. The misidentification probability of electrons and
muons is about 1\%. This high efficiency makes it possible to study details
of CC $\nu_e$ and $\nu_{\mu}$ events.  Figure~\ref{fig:zenith} shows the
zenith angle distributions for contained atmospheric neutrino events. An
energy and zenith angle dependent deficit of $\mu$-like events is clearly
visible. From these distributions, it was concluded that the observed data
show a compelling evidence for neutrino oscillations \cite{sk-evidence}.

\begin{figure}
\includegraphics[width=3.1in]{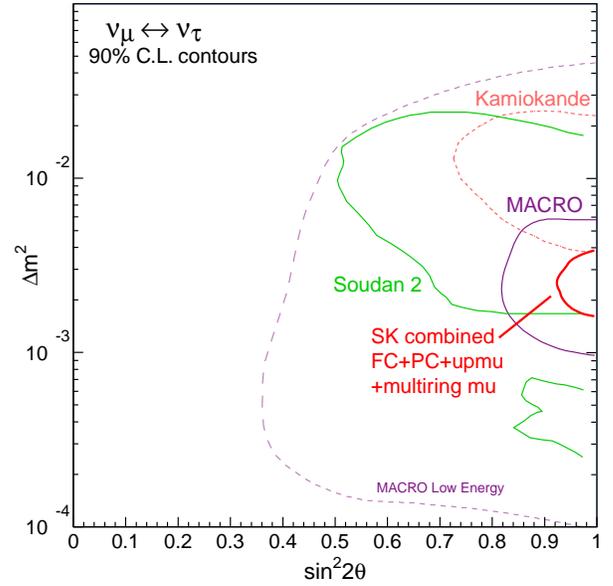}
\caption{\label{fig:atm-osc}
90\% C.L. allowed parameter regions of oscillation parameters from
Super-Kamiokande(thick line), Kamiokande(thick broken line),
Soudan-2(thin line) and MACRO(thin broken line).
Two flavor  $\nu_{\mu} \rightarrow \nu_{\tau}$ oscillations
are assumed. }
\end{figure}

Since the zenith angle distributions for $e$-like events show no evidence for
inconsistency between data and non-oscillated Monte Carlo, it can be
concluded that the oscillation could be between $\nu_{\mu}$ and $\nu_{\tau}$.
Indeed, detailed oscillation analyses of the Super-Kamiokande
data \cite{sk-evidence,sk-sterile} have concluded that the data are completely
consistent with two-flavor $\nu_{\mu} \rightarrow \nu_{\tau}$ oscillations.
The present understanding of the oscillation parameters ($sin ^2 2 \theta,~
\Delta m^2 $) for $\nu_{\mu} \rightarrow \nu_{\tau}$ two flavor oscillation
is shown in Figure~\ref{fig:atm-osc}. Results from 4 experiments are shown,
and all the results are essentially consistent.  The most accurate result
from Super-Kamiokande gives oscillation parameters of: ${\rm sin}^2 2 \theta
> 0.92$ and $1.6 \times 10^{-3} < \Delta m^2 < 3.9 \times 10^{-3}$eV$^2$ at
90\% C.L.

\begin{figure}
\includegraphics[width=3.1in]{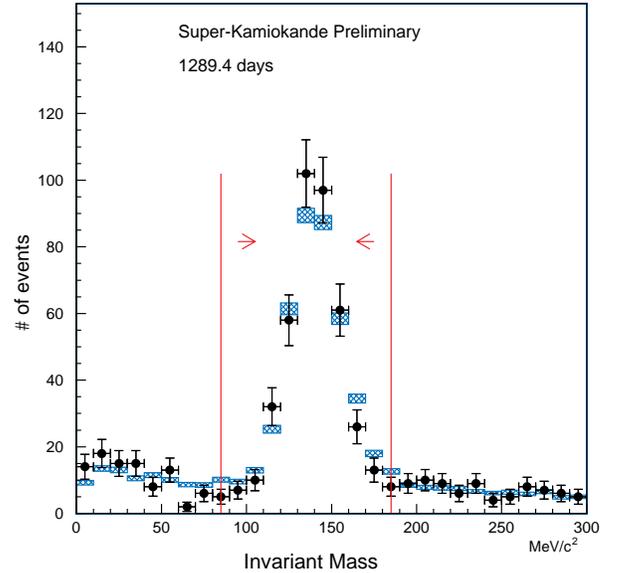}
\caption{\label{fig:pi0}
Invariant mass distribution for two electron-like ring events 
observed in Super-Kamiokande based on 79 ktonyr data. Data are shown as filled circles,
 simulation as hatched rectangles.}
\end{figure}

In addition, it is possible for water Cherenkov detectors
to study more complicated events. An example of such 
events are those containing a single $\pi^0$. These events are characteristic of neutral
 current neutrino interactions, and are therefore useful in the study of neutrino
 oscillations to sterile neutrinos. Since the $\pi^0$ decays to two
gamma rays, these events are comprised of two $e$-like ring
events.  Figure~\ref{fig:pi0} shows  the invariant mass distribution for two
electron-like events.  A clear peak is seen at the $\pi^0$ mass.

It is possible to construct a water Cherenkov detector larger than
Super-Kamiokande. It is widely thought that the next generation water
Cherenkov detectors should have the mass of about 1~Mton, in order to carry
out large statistics neutrino oscillation studies with neutrino beams, and in
order to probe proton decay lifetimes approaching $10^{35}$ years, as
predicted by modern Grand Unified Theories. These detectors
would also observe significantly larger number of atmospheric neutrino
events. By exploiting the high statistics and the up-down symmetry of the
neutrino flux, it should be possible to determine sin$^2 2\theta$ to an
accuracy of about 1\%. This is possible, because $N_{up}/N_{down} = 1 -
\sin^2 2 \theta / 2$ to first approximation, where $N_{up}$ and $N_{down}$
show the number of upward going and downward going events.  Since the
dimension of the detector is larger, muons with energy higher than 10~GeV
will be contained in the detector.  It could then be possible to observe the
``oscillation'' nature of the $\nu_{\mu}$ disappearance that has not been
observed by any oscillation experiment. In addition, a larger detector with
good $\nu_e$ reconstruction could observe the yet-unobserved mixing angle
($\theta_{13}$) through the Earth-matter effect, if the mixing angle is near
the present limit~\cite{chooz}.

One of the technical requirements for such ultra-large water Cherenkov
detectors was exposed in 2001 by the accidental loss of photomultiplier tubes
by a chain reaction caused by a single imploding tube in the Super-Kamiokande
detector. Detailed studies were subsequently performed, both with
hydrodynamic computer simulations and in-situ tests under 4.5 atmospheres of
water pressure. Based on these studies, a two-piece protective shell was
designed, consisting of $\sim 8$ mm fibre reinforced plastic and 13 mm clear acrylic.  The shell
is not intended to withstand static pressure, but has several small holes
which allow water to form a layer between the PMT and the shell. If the PMT
breaks, water rushes in through the holes relatively slowly, and does not
produce the sharp pressure wave that caused the 2001 accident. The loss of
light due to the extra water-acrylic interfaces is estimated to be a few
percent and will be properly modelled in the detector simulation.

\subsection{Fine Grained Tracking Detectors}

\begin{figure}
\includegraphics[width=3.3in]{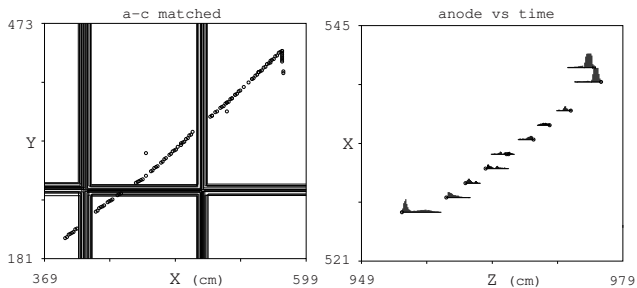}
\caption{\label{fig:s2-event}
  An example event display from the Soudan detector, showing the long
  track from a muon and a shorter, more heavily ionizing track from a
  recoil proton.}
\end{figure}

\begin{figure}
\includegraphics[width=3.1in]{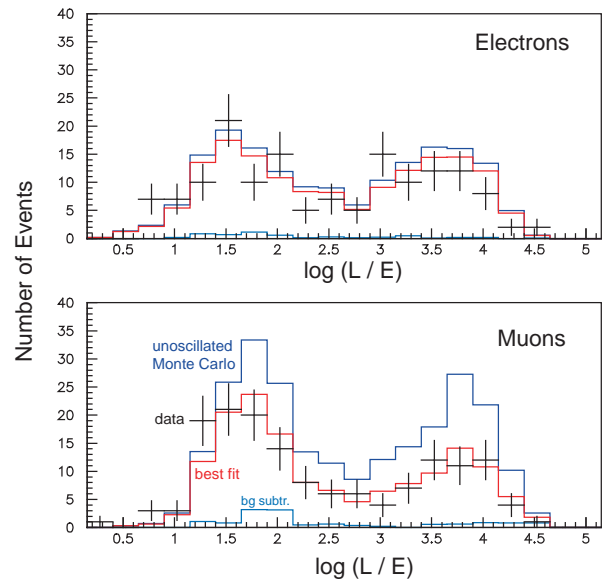}
\caption{\label{fig:s2lovere}
The $L/E$ distribution for high resolution atmospheric neutrinos in the
  Soudan 2 analysis.}
\end{figure}

The second major category of atmospheric neutrino detectors consists of
comparatively fine resolution tracking detectors. The first generation of
these experiments includes KGF \cite{kgf}, NUSEX \cite{nusex}, Frejus \cite{frejus-detector} and
Soudan~1. The Soudan~1
detector was 30 tons of iron-loaded concrete instrumented with proportional
tubes. The design, but not the site, was changed for the later Soudan 2
detector~\cite{soudan}.  These first generation
detectors were smaller than the water Cherenkov detectors, but in most cases
became operational somewhat earlier. Tracking detectors have an
advantage in sensitivity because they can detect low velocity charged
particles that would be below Cherenkov threshold in water. In particular,
the Soudan 2 detector is able to reconstruct the short and heavily ionizing
trajectory of recoil protons from atmospheric neutrino events such as
$\nu_\mu + n \rightarrow \mu^- + p$, as shown in Fig.~\ref{fig:s2-event}.
The complete specification of the two-body final state allows for the
selection of low energy neutrino events with well-measured direction and
energy, which is important to neutrino oscillation through the ratio $L/E$.
The Soudan collaboration combines these quasi-elastic events with a higher
energy sample including events with visible energy greater than 600 MeV and
with multiple tracks in the final state. They estimate a pointing resolution
of $20^\circ-30^\circ$ and a ${\rm log} (L/E) $ resolution of 0.5. Analysis of
this data, shown in Fig.~\ref{fig:s2lovere} yields the confidence interval
shown in Fig.~\ref{fig:atm-osc}.

Liquid argon TPCs such as the upcoming ICARUS experiment \cite{icarus} will
reconstruct event details including the ionization level of charged tracks
with unprecedented resolution. The first generation detectors will be
relatively small; the first installation of ICARUS at Gran Sasso will be 600
tons, with a planned increase to 3 ktons. There are ideas to build even
larger liquid argon TPCs~\cite{lanndd}.

\begin{figure}
\includegraphics[width=3.1in]{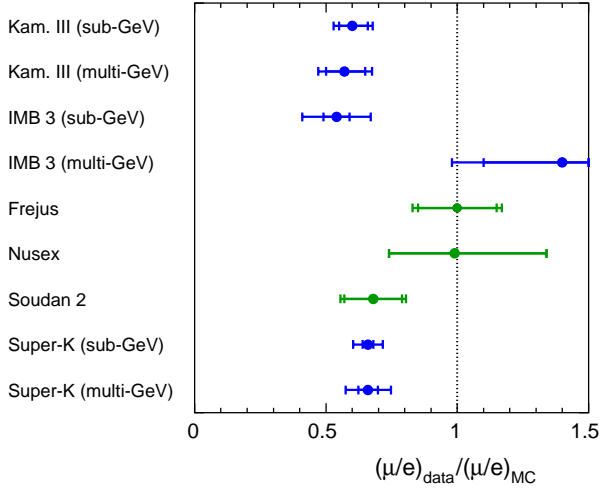}
\caption{\label{fig:RvsExp}
A compendium of the measurements of the double ratio $\nu_\mu/\nu_e$(data)
to $\nu_\mu/\nu_e$(M.C.) for atmospheric neutrino experiments.}
\end{figure}

No significant signal of proton decay was ever detected by any of the
iron-based tracking experiments, however each did measure the background from
atmospheric $\nu_\mu$ and $\nu_e$ interactions. In the early history of the
atmospheric neutrino anomaly, a dichotomy appeared where the water Cherenkov
detectors Kamiokande~\cite{kamR} and IMB \cite{IMBR} measured a $\nu_\mu/\nu_e$
ratio inconsistent with expectation, whereas the non-water detectors
NUSEX \cite{NUSEXR} and Frejus \cite{FrejusR} did not, although with poorer
event statistics. This is shown in Fig.~\ref{fig:RvsExp}, including data
points from more recent experiments as well. The difference between water and
iron detectors led to studies of neutrino interactions with light nuclei
versus heavy nuclei, but resulting in no identification of a cause for the
different results. Ultimately, the explanation is likely to be statistical,
as the larger data set from Soudan 2 now clearly measures an anomalous
$\nu_\mu/\nu_e$ ratio with the hallmarks of neutrino
oscillation \cite{SoudanR}.

\begin{figure}
\includegraphics[width=3.1in]{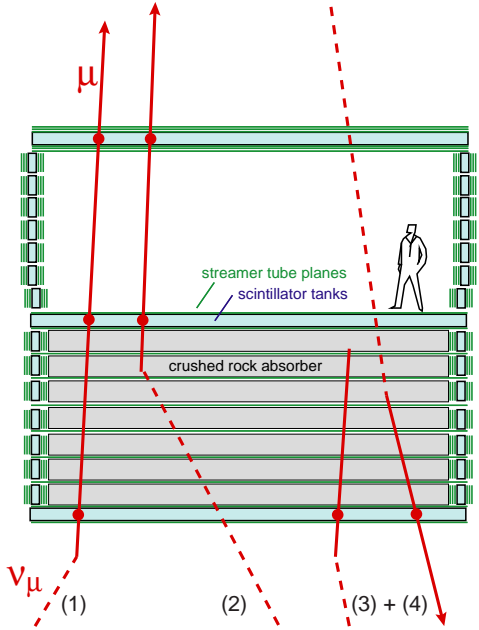}
\caption{\label{fig:macrodet}
An end view of the MACRO detector, showing different categories of
neutrino-induced events: (1) upward through-going muon events, (2) upward partially-contained events,
(3) upward stopping muon events, and (4) downward partially contained events.}
\end{figure}

There is a second category of fine grained tracking detector that is mostly
sensitive to muon neutrinos in the form of upward-going muons. These include
the Baksan and MACRO experiments, which have fairly small absorber mass (a
few hundred tons) and were never intended to detect proton decay. 
The MACRO detector, shown in Fig.~\ref{fig:macrodet}, was composed of three horizontal
planes of liquid scintillator tanks separated by several meters, with the
lower section filled with crushed rock absorber and a hollow upper ``attico''
section. The sides of the detector are also covered with vertical planes of
scintillator.  Analysis of the upward through-going muon sample yields the
confidence interval shown in Fig.~\ref{fig:atm-osc}. In addition to
through-going muons, MACRO has analyzed partially contained and stopping
topologies, where the crushed rock in the lower section acts as target or
stopping absorber respectively \cite{macropcstop}.

\begin{figure}
\includegraphics[width=3.1in]{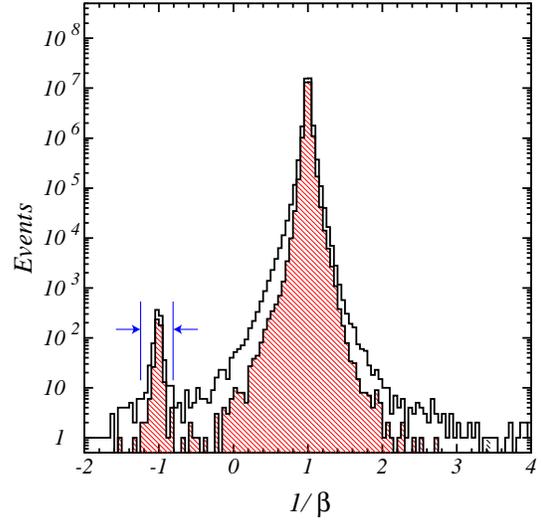}
\caption{\label{fig:macrobeta}
Distribution of $1/\beta$ for all muons in the data.
A clear peak of upward-going muons is evident. The shaded
part of the distribution is for the subset of events
where three scintillator layers were hit.}
\end{figure}

Time-of-flight detectors identify the direction of the muon by measuring the
time interval as the muon traverses two or more layers of liquid scintillator.
Figure~\ref{fig:macrobeta} shows the $1/\beta$ distribution for throughgoing
upward muons reconstructed by the MACRO experiment, clearly separating
several hundred neutrino induced events from the background of several
million downward-going cosmic rays \cite{macroupmu}.

Fine-grained tracking also allows a crude muon energy estimate by measuring
the amount of multiple coulomb scattering (MCS). One of several empirical
formula \cite{pdg} gives the r.m.s. lateral displacement of a relativistic
muon after crossing absorber thickness $X$, measured in radiation lengths:

\begin{equation}
\sigma^{MCS}_{proj} \sim \frac{X}{\sqrt{3}} \frac{13.6 MeV}{p_\mu \beta c}\sqrt{\frac{X}{X_0}}(1+0.038 \ln \frac{X}{X_0}).
\end{equation} \noindent

For MACRO, $\sigma^{MCS}_{proj} \sim 10 cm/E_\mu(GeV)$. Because the MACRO
streamer tube tracking was designed to measure the transit of slow particles
such as magnetic monopoles, the detector is equipped with timing electronics
that records the streamer drift times. This allows a spatial resolution of
$\sim 0.3$ cm (in a 2.9 cm wide cell), verified by beam test
studies. The technique saturates for muon energies above
40 GeV, below which the parent neutrino energy may be estimated with a
resolution of $\delta E_\nu/E_\nu \sim 150\%$.

There is a final category of atmospheric neutrino detector that includes
magnetic tracking. The first such detector, MINOS \cite{minos}, was designed
mostly for a long-baseline accelerator beam, and therefore has a built in
orientation along the neutrino beam direction. MINOS will be composed of 486
layers of iron sandwiched with 4-cm wide by 1-cm thick strips of plastic
scintillator, with a total mass of 5.4 ktons. The average magnetic field
strength is 1.3 Tesla. MINOS can make a unique contribution to atmospheric
neutrino studies by measuring the $\nu_\mu/\bar{\nu}_\mu$ ratio. This will
test the near unity ratio predicted by atmospheric neutrino flux models, and
possibly provide CPT tests of different oscillation probability for $\nu_\mu$
and ${\bar \nu}_\mu$.

In addition, the magnetic field of MINOS will allow for the momentum
determination of exiting muons up to approximately 70 GeV. This will provide
a sample of $\nu_\mu$ interactions with relatively accurate $L$ and $E_\nu$
determination for neutrino oscillation studies.  This analysis was also
proposed for a much larger underground neutrino detector called
MONOLITH \cite{monolith}. MONOLITH would use a very large mass of magnetized
iron for the goal of measuring $L/E$ with resolution sufficient to resolve
the oscillatory pattern in atmospheric neutrino oscillations \cite{monolith}.

\section{High Energy Neutrino Telescopes}

\begin{figure}
\centering
\includegraphics[width=6cm]{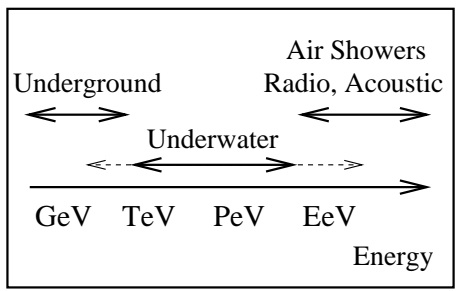}
\caption{\label{methodscs}
Energy range of the various detection techniques (see below). Optical 
Cherenkov detectors, although optimized to the TeV-PeV range, are sensitive 
also at lower and higher 
energies, as indicated by the dashed lines.
}
\end{figure}

Whereas MeV neutrino astronomy has been established by 
the observation  of solar neutrinos 
and neutrinos from supernova SN1987A,  
neutrinos with energies of GeV to PeV ($10^6$ GeV), 
which must accompany the production of  high  energy 
cosmic rays, still await discovery. 
Detectors underground have turned out to be too small to detect 
the feeble fluxes of  energetic neutrinos from cosmic accelerators. 
The high energy frontier of TeV (= $10^3$ GeV)
and PeV  energy is currently being tackled by much larger, 
expandable arrays constructed 
in open water or ice. Detectors tailored to record acoustic, radio, 
fluorescence or air 
shower signatures from neutrino interactions at EeV energy 
(= $10^9$ GeV) and above 
are being designed in parallel. 
Fig.~\ref{methodscs} sketches the energy domains of different techniques.

\subsection{Physics Goals}

The central goal of high energy neutrino telescopes is to settle 
the origin of high  energy 
cosmic rays \cite{LM}. The directional information of these 
charged particles - protons, light and heavy nuclei -  is lost 
due to deflection in cosmic magnetic fields 
(apart from the extreme energies above $10^{10}$ GeV where 
deflection is negligible). Source 
tracing, i.e. {\it astronomy}, is only possible by neutral, stable 
particles like $\gamma$ rays and neutrinos. 
In contrast 
to $\gamma$ rays which may come from pure electron acceleration, 
only neutrinos provide incontrovertible evidence of proton acceleration. 
On top of that, neutrinos do not suffer from absorption by the omnipresent 
infrared or radio background when propagating through space. 
The range of TeV $\gamma$ rays is only about 100 Mpc, 
at PeV only 10 kpc, i.e. about the radius of our Galaxy. 
Therefore, the topology of the far distant high energy Universe may  
possibly be investigated only with neutrinos.

The physics goals of high energy neutrino telescopes include:

\begin{enumerate}
\item [a)] Search for neutrinos from cosmic acceleration processes in galactic 
sources like micro quasars or supernova remnants (SNR), 
or extragalactic sources like active galactic nuclei (AGN) or 
gamma ray bursts (GRB),
\item [b)] search for ultra-high energy (UHE) neutrinos
from interactions of UHE cosmic rays with the photons of 
the cosmic 3K microwave background (the so called GZK --
Greisen-Zatsepin-Kuzmin -- neutrinos),
from topological defects (TD) or 
from the decay of super-heavy particles,
\item [c)] search for neutrinos from the annihilation of 
Weakly Interacting Massive Particles (WIMPs),
\item [d)] search for magnetic monopoles,
\item [e)] monitoring our Galaxy for MeV neutrinos from supernova bursts.\\
\end{enumerate}

Most models related to sources of type 
{\it a)} assume acceleration by shock waves propagating in 
accretion discs around black holes or along the extended jets emitted 
perpendicularly to the disk (bottom-up models). 
Neutrinos are generated in decays of mesons produced by interactions of 
the accelerated charged particles with ambient matter or with photon gas. 

\begin{eqnarray*}
p + p (\gamma) \rightarrow   p (n) +  \pi\\
                             &  \searrow  \mu +  \nu
\end{eqnarray*}

The neutrino energy spectrum of many models follows an 
$E_{\nu}^{-2}$ behaviour, 
at least over a certain range of energy. 
Assuming an $E_{\nu}^{-2}$ form and normalizing the neutrino flux 
to the measured 
flux of cosmic rays at highest energies leads to an upper bound of 
$dN/dE_{\nu} \sim 5 \times 10^{-8} E_{\nu}^{-2}$ GeV$^{-1}$ cm$^{-2}$ 
s$^{-1}$ sr$^{-1}$ to the diffuse 
neutrino flux (i.e. the flux integrated over all possible sources) \cite{WB}.
Reasonably weakened assumptions loosen this bound by more than one 
order of magnitude to 
$10^{-6} E_{\nu}^{-2}$ GeV$^{-1}$ cm$^{-2}$ s$^{-1}$ sr$^{-1}$ ~\cite{MPR}(see also Fig.\ref{diffuse}). 
The so-called top-down scenarios of 
type {\it b)} are suggestive for the explanation of highest 
energy cosmic rays. 
In this scenario, high energy particles would be ``born'' 
with high energies, and not accelerated from low to high energies, 
as in the standard bottom-up scenarios. 

We will focus on {\it a)} and {\it b)} in the following and refer to ~\cite{LM,GHS,CS2000,CS2002} 
and references therein for more information on {\it c)} - {\it e)}.

\subsection{Cherenkov telescopes under water and ice}

Optical underwater/ice neutrino detectors consist of a lattice 
of  photomultipliers (PMTs) housed in transparent pressure spheres which are  
spread over a large open volume  in the ocean, in lakes or in ice. 
In most designs the spheres are attached to strings which - in the 
case of water detectors - are moored at the ground and held vertically 
by buoys. The typical spacing along a string is 10-20 meters, 
and between strings 30-100 meters. The spacing is very large 
compared to Super-Kamiokande. This allows large volumes to be covered but 
makes the detector practically blind with respect to phenomena below 10~GeV. 

\begin{figure}
\centering
\includegraphics[width=8.3cm]{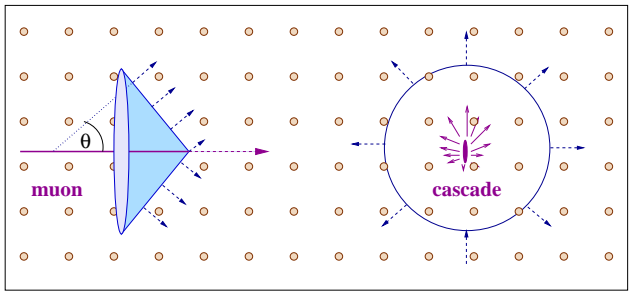}
\caption{\label{detmodes} 
Detection of muon tracks (left) and cascades (right).
}
\end{figure}

The PMTs record arrival time and amplitude of Cherenkov light emitted by 
muons or particle cascades. The accuracy in time is a few nanoseconds. 
Fig.~\ref{detmodes} sketches the two basic detection modes.
 
In the {\it muon} mode, high energy neutrinos are inferred from 
the Cherenkov cone accompanying muons which enter the detector from below. 
Such upward moving muons can have been produced only in interactions 
of muon neutrinos having crossed the earth. 
The effective volume considerably exceeds the actual detector volume 
due to the large range of muons (about 1 km at 
300 GeV and 24 km at 1 PeV). 
Muons which have been 
generated in the earth atmosphere above the detector and 
punch through the water or ice down to the detector, outnumber
neutrino-induced upward moving muons by several
orders of magnitude and have to be removed by 
careful up/down assignment. 
At energies above a few hundred TeV, where the earth is going to become 
opaque even to $\nu_e$ and $\nu_\mu$ neutrinos, muons arrive only from directions close to the horizon, 
at EeV energies even only from the upper hemisphere. 
Most of these muons can be distinguished from down going atmospheric 
muons due to their higher energy deposition.

Apart from elongated tracks, {\it cascades} can be detected. 
Their length increases only like the logarithm of the cascade energy. 
With typically 5-10 meters length, and a diameter of the order of 10 cm,  
cascades may be considered as quasi point-like compared to the spacing of 
photomultipliers in Cherenkov telescopes. The effective volume  
for cascade detection is close to the geometrical volume. 
While for present telescopes it therefore is much smaller than that 
for muon detection, for kilometer-scale detectors and not too large 
energies it can reach the same order of magnitude as the latter.

Underwater/ice telescopes are optimized for the detection of muon tracks 
and for energies of a TeV or above, by the following reasons: 

\begin{enumerate}
\item [a)] Above one TeV, the spectrum of atmospheric neutrinos
changes from $E^{-3}$ to $E^{-3.7}$ dependence.
This is significantly steeper than the expected
spectrum of neutrinos from cosmic accelerators 
and results in 
a much better signal-to-background ratio at higher energies.
\item [b)] Neutrino cross section and muon range increase with energy.  
The larger the muon range, the larger is the effective detection volume.
\item [c)] The mean angle between muon and neutrino decreases with energy 
like $E^{-0.5}$, with a pointing accuracy of about one degree at 1 TeV.
\item [d)] Mainly due to 
pair production and bremsstrahlung, the energy loss of 
muons increases with energy. For energies above 1 TeV, this allows an 
estimate of the muon energy from the larger light emission along the track.
\end{enumerate}

\begin{figure}
\centering
\includegraphics[width=3.6cm]{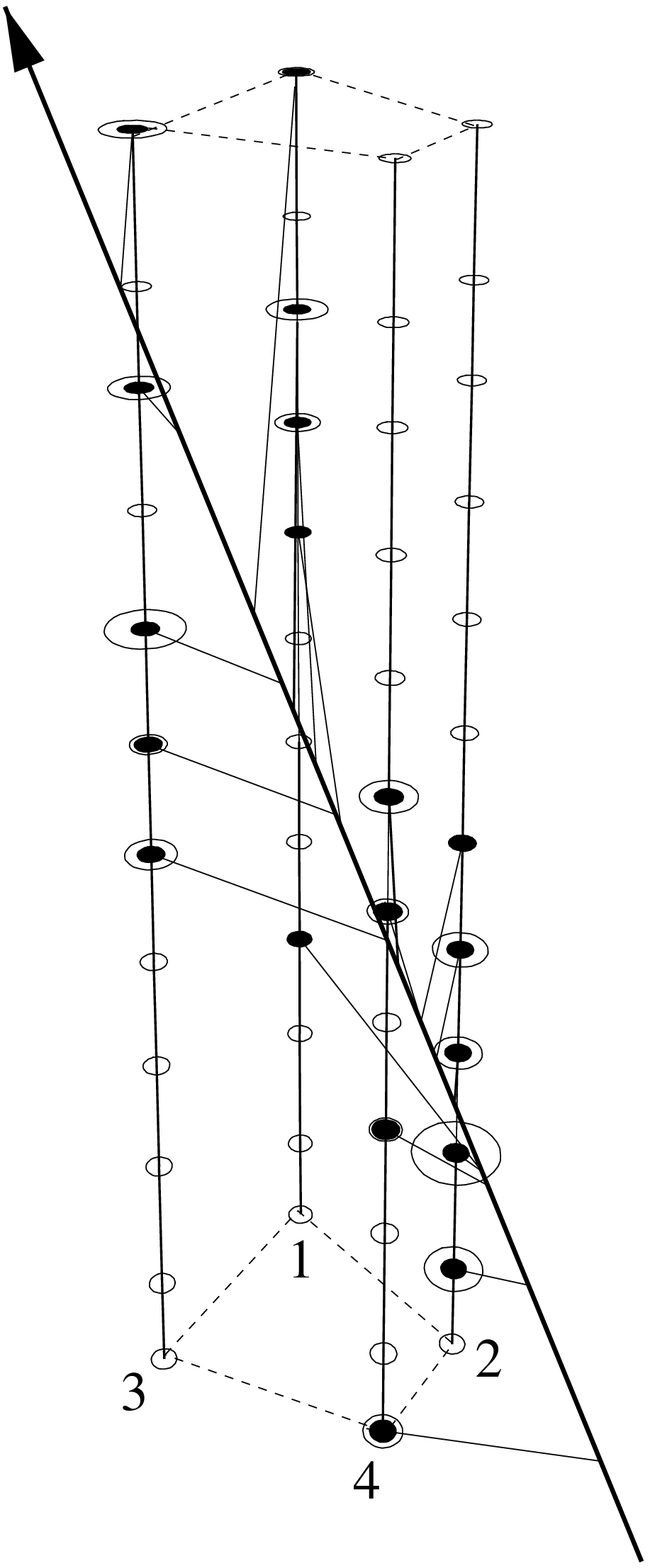}
\hspace{0.7cm}
\includegraphics[width=3.6cm]{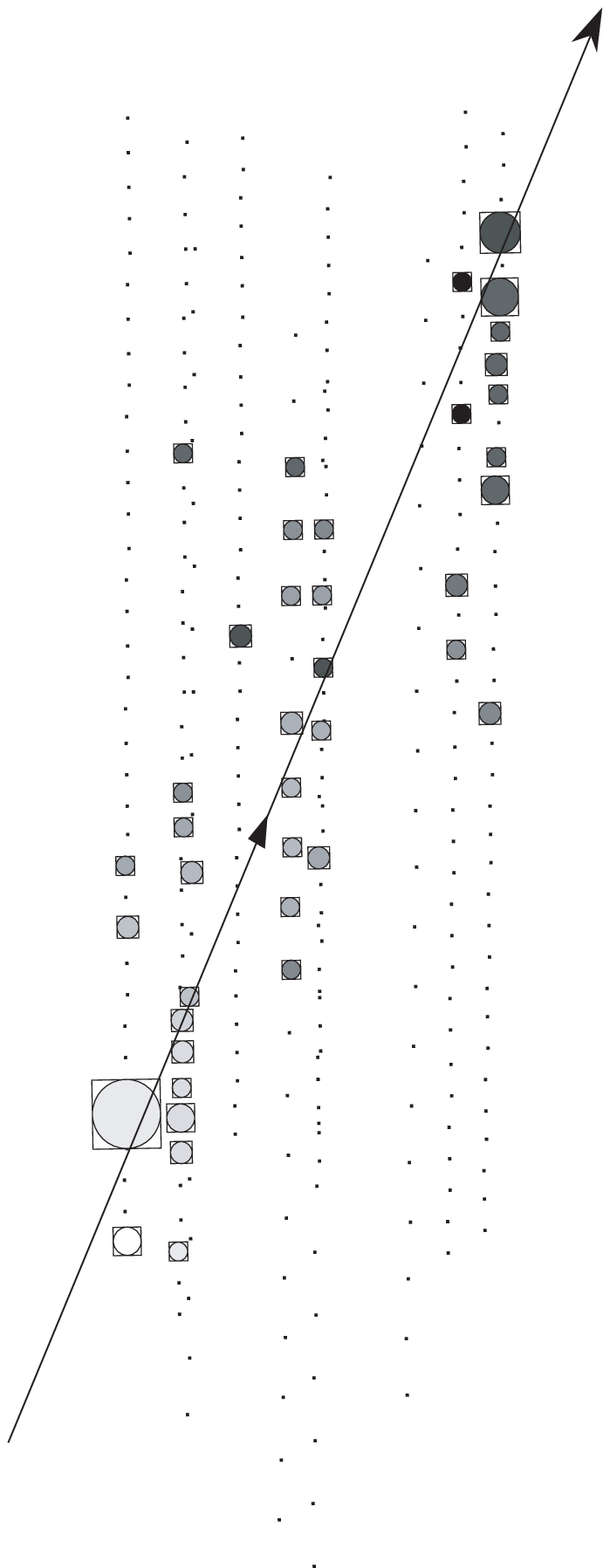}
\caption{\label{nuevents} 
Left: one of the first clearly upward moving muons recorded with the 1996 
four-string-stage of the Baikal detector. Small ellipses denote PMTs. 
Hit PMTs are black, with the size of the disc proportional to the 
recorded amplitude. The arrow line represents the reconstructed muon track,
the thin lines the photon paths.
Right: Upward muon recorded by the 1997 version of AMANDA. 
Small dots denote the PMTs arranged on ten strings.  
Hit PMTs are highlighted by boxes,
with the degree of shadowing indicating the time (dark being late),
and the size of the symbols the measured amplitude.
Note the different scales: the height of the Baikal array is 72 
meters, that of AMANDA nearly 500 meters.
}
\end{figure}

The development in this field was stimulated originally by the DUMAND project 
close to Hawaii which was cancelled in 1995. The breakthrough came from 
the other pioneering experiment located at a depth of 1100 m in the 
Siberian Lake Baikal. The BAIKAL collaboration not only 
was the first to deploy three strings (as necessary for full spatial 
reconstruction \cite{BAIKAL1}), but also reported the first atmospheric neutrinos 
detected underwater (\cite{BAIKAL}, see Fig.~\ref{nuevents},\,left). 
At present, NT-200 is taking data, an array comprising 192  mushroom 
shaped 15''-PMTs on 4 strings. 
A moderate upgrade (NT200+) is planned for 2004.

\begin{figure}
\centering
\includegraphics[width=8.2cm]{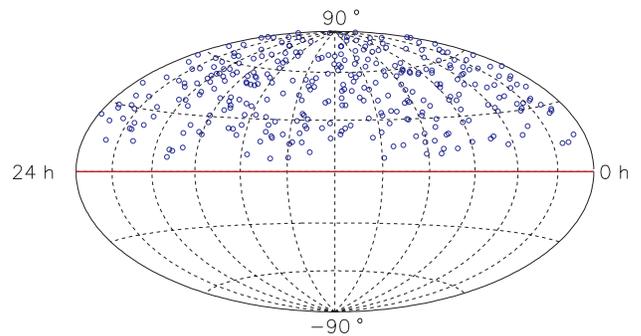}
\caption{\label{skymap} 
Sky map of 300 neutrino candidates taken with AMANDA B10 in 1997. 
No indication of 
extraterrestrial point sources on top of  atmospheric neutrinos are found.
}
\end{figure}

With respect to its size, NT-200 has been 
surpassed by the AMANDA detector \cite{AMANDA}.  
Rather than water, AMANDA uses the 3 km thick ice layer at the 
geographical South Pole. 
Holes are drilled with hot water, 
and strings with PMTs are frozen into the ice. 
With 677 PMTs at 19 strings, most at depths between 1500-2000 m, 
the present AMANDA-II array reaches an area of a few 10$^4$ m$^2$ for 
1 TeV muons. AMANDA-II may be the first detector 
with a realistic discovery potential with respect to extraterrestrial high 
energy neutrinos, even though it is smaller than the square kilometer size generally
predicted to be required for clear observation of such signals. 
Limits obtained from the analysis of data taken with the 
smaller ten-string detector AMANDA-B10 in 1997 are similar to or below those 
limits which have been obtained by underground detectors over more than a 
decade of data taking. The limit \cite{Hill} on the diffuse flux from unresolved sources 
with an assumed $E^{-2}$ spectrum is 
$0.8 \cdot 10^{-6} E_{\nu}^{-2}$ GeV$^{-1}$ cm$^{-2}$ s$^{-1}$ sr$^{-1}$,  
slightly below the loosest theoretical bounds \cite{MPR} and a factor of 2 
below the corresponding Baikal limit.  AMANDA limits on point sources 
on the Northern sky \cite{point} complement the limits obtained 
from detectors in the 
Northern hemisphere for the Southern sky (see Fig.~\ref{point}). 
The overall sensitivity of AMANDA-B10 has been verified by samples of 
events which are dominated by atmospheric neutrinos \cite{Amanda}.  
Fig.~\ref{nuevents}\,(right)  shows an neutrino event taken 
with AMANDA-B10, Fig.~\ref{skymap} 
shows the sky map of the first 300 neutrino candidates taken in 1997.

Based on the experience from AMANDA, a cubic kilometer detector, 
ICECUBE \cite{ICECUBE}, is going to be deployed at the South Pole. 
It will consist of 4800 PMTs on 80 vertical strings, 
with 125 m inter-string-distances and a 16 m spacing between 
the PMTs along a string. The 8-inch AMANDA PMTs will be replaced by 
10-inch PMTs, and the simple analog-readout by a digital readout of 
the full transient waveform recorded by the PMT.

Two projects for large neutrino telescopes are under construction 
in the Mediterranean - ANTARES \cite{ANTARES} and  
NESTOR \cite{NESTOR} .  
Both have assessed the relevant physical and optical 
parameters of their sites and have deployed prototype arrays of 
about a dozen PMTs. 
ANTARES and NESTOR follow different deployment schemes 
and array designs. The NESTOR group plans to deploy a tower of 
several floors, each 
carrying 12 PMTs at 16 m long arms. The ANTARES detector will consist 
of 12 strings, 
each equipped with 30 triplets of PMTs. 
This detector will have an area of 
about 2 $\cdot$ 10$^4$ m$^2$ for 
1 TeV  muons - similar to AMANDA-II - and is planned to 
be fully deployed in 2006. 
An additional initiative, NEMO, 
has finished a series of site explorations at a location  
70 km from Sicily and is now in the 
phase of prototype studies for a cubic kilometer 
detector \cite{NEMO}. The different efforts are going to converge towards one single 
project for a cubic kilometer detector.

There have been longstanding discussions about the best location for  
a future large neutrino telescope. One detector in each hemisphere will be necessary for full sky coverage. 
With respect to optical properties, water detectors in oceans seem 
to be favoured: although the absorption length  of  Antarctic ice at 
Amanda depths is more than 1.5 times that in oceans 
(and about four times that of Baikal), ice is characterized by strong 
light scattering, and its optical parameters vary with depth. 
Light scattering leads  to a considerable delay of Cherenkov photons. 
On the other hand ice does not suffer from the high potassium content 
of ocean water or from bioluminescence. These external light sources result 
in counting rates ranging from several tens of kHz to a few hundred kHz per PMT, 
compared  to less than 500 Hz per PMT dark count rate in ice.  
Depth arguments favour oceans.  
Note, however, that this argument lost some of its 
initial  strength after BAIKAL and AMANDA developed reconstruction 
methods which effectively reject even the high background at shallow depths. 
What counts most, at the end, are basic technical questions like deployment, 
or the reliability of the single components as well as of the whole system. 
Systems with a non-hierarchical structure like AMANDA (where each PMT has 
its own  2 km cable to surface) will  suffer less from single point 
failures than water detectors do.  In the case of water,  longer distances 
between the detector and the shore station have to be bridged. 
Consequently, not every PMT can get its own cable to shore, 
resulting in a hierarchical system architecture. 
This drawback of water detectors may be balanced by the fact that they 
allow retrieval and replacements of failed components, as the BAIKAL 
group has demonstrated over many years.

\subsection{Acoustic detection}

A high energy particle cascade deposits energy into the medium via ionization losses, 
which is immediately converted into heat. 
The effect is a  fast expansion, generating a bipolar acoustic pulse. 
Transverse to the pencil-like cascade (diameter about 10 cm)  
the radiation 
propagates within a disk of about 10 m 
thickness (the length of the cascade) into the medium. The 
signal power peaks 
at 20 kHz where the attenuation length of sea water is a few kilometers, 
compared to a few tens of meters for light. Given a large initial signal, 
huge detection volumes can be achieved. With efficient noise rejection, 
acoustic detection might be competitive with optical  
detection at multi-PeV energies \cite{Acoustics} .
  
Present initiatives \cite{CS2002} envisage combinations of 
acoustic arrays with  optical 
Cherenkov detectors (NESTOR, ANTARES, ICECUBE) or the use of 
existing sonar arrays for submarine detection close to Kamchatka and in 
the Black Sea \cite{Capone}. Most advanced is AUTEC,
 a project using a very large hydrophone array of the US Navy, 
close to the Bahamas \cite{AUTEC} .  
The existing array of 52 hydrophones spans an 
area of 250 km$^2$ and has good sensitivity 
between 1-50 kHz. Due to the sparse instrumentation,
it is expected to trigger only on events above 100~EeV.

\subsection{Radio detection}

Electromagnetic showers generated by high energy electron neutrino 
interactions emit coherent  Cherenkov radiation.
Electrons are swept into 
the developing shower, which acquires a negative net charge from the added 
shell electrons. This charge propagates like a relativistic pancake of 1 cm 
thickness and 10 cm diameter. Each particle emits Cherenkov radiation, 
with the total signal 
being the resultant of the overlapping Cherenkov cones. 
For wavelengths larger than the cascade diameter,  
coherence is observed and the signal rises proportional to $E^2$, 
making the method attractive for high energy cascades. The bipolar radio 
pulse has a width of 1-2 ns.  In ice as well as in salt domes, 
attenuation lengths of several kilometers can be obtained, depending 
on the frequency band, the temperature of the ice, and the salt quality. 
Thus, for energies above a few tens of PeV, radio detection in ice 
or salt might 
be competitive or superior to optical detection.

A prototype Cherenkov radio detector called RICE is operating at the 
geographical  South Pole \cite{RICE}. Twenty receivers and emitters are buried at depths 
between 120 and 300 m. 
From the non-observation of very large pulses, a limit of about 
$10^{-5}  E_{\nu}^{-2}$ GeV$^{-1}$ cm$^{-2}$ s$^{-1}$ sr$^{-1}$ 
has been derived for 
energies above 100 PeV \cite{Krav}. SALSA, a R\&D project study for radio 
detection in natural salt domes, 
promises to get a limit about three orders 
of magnitude better \cite{SALSA} . 
ANITA (ANtarctic Impulsive Transient Array)  
is an array of radio antennas planned to be 
flown at a balloon on an Antarctic circumpolar path in 2006 \cite{ANITA}. 
From 35 km altitude it may 
record the radio pulses from neutrino interactions in the 
thick ice cover and monitor a 
really huge volume. 
The expected sensitivity from a 30 day flight is about 
$10^{-7} E_{\nu}^{-2}$ GeV$^{-1}$ cm$^{-2}$ s$^{-1}$ sr$^{-1}$ 
at 10 EeV.

Most exotic is the Goldstone Lunar Ultrahigh Energy Neutrino 
Experiment, GLUE. 
It has searched for radio emission from extremely-high energy 
cascades induced by neutrinos 
or cosmic rays skimming the moon surface \cite{GLUE}. 
Using two NASA antennas, 
an upper limit of 
$5 \cdot 10^{-5} E_{\nu}^{-2}$ GeV$^{-1}$ cm$^{-2}$ s$^{-1}$ sr$^{-1}$ 
at 100 EeV has been obtained.

\subsection{Detection of neutrino energies via air showers}

At supra-EeV energies, large extensive air shower arrays like the
AUGER detector in Argentina \cite{AUGER} or the
telescope array \cite{TA}
may search for horizontal air showers due to neutrino interactions deep in 
the atmosphere (showers 
induced by charged cosmic rays start on top of the atmosphere). 
AUGER consists of an array of water tanks that will 
span an area of more than 3000 km$^2$ and 
will record the Cherenkov light of air-shower particles crossing the tanks. 
It is combined with telescopes looking for the atmospheric fluorescence 
light from air showers. The optimum sensitivity window for this 
method is at 1-100 EeV, the effective detector mass is between 1 and 20 Giga-tons, 
and the estimated sensitivity is of the order of  
$10^{-7} E_{\nu}^{-2}$  GeV$^{-1}$ cm$^{-2}$ s$^{-1}$ sr$^{-1}$. 
An even better sensitivity might be obtained for tau neutrinos, $\nu_{\tau}$, 
scratching the Earth and interacting close to the array.
The charged $\tau$ lepton produced in the interaction can escape 
the rock around the array,  in contrast to electrons,
and in contrast to muons it decays after a 
short path into hadrons \cite{Fargion}. 
If this decay happens above the array or in the field of view 
of the fluorescence telescopes, the decay cascade can be recorded. 
Provided the experimental pattern allows clear identification,  
the acceptance for this kind of signals can be large. 
For the optimal energy scale of 1 EeV, the sensitivity might reach  
$10^{-8} E_{\nu}^{-2}$ GeV$^{-1}$ cm$^{-2}$ s$^{-1}$ sr$^{-1}$.
A variation of this idea is to search for tau lepton cascades which are
produced by horizontal PeV neutrinos hitting a  mountain and then
decay in a valley between target mountain and an ``observer'' mountain \cite{mountain}.

Already in the middle of 1990's, 
the Fly's Eye collaboration \cite{Fly},
and more recently,
the Japanese AGASA collaboration \cite{AGASA}
have practiced the search mode of horizontal 
air showers. AGASA derived an upper limit of the order of 10$^{-5}$ 
in the units given above 
- only just one order of magnitude above some 
predictions for AGN jets and for topological defects.

Heading to higher energies leads to space based detectors 
monitoring  larger volumes than 
visible from any point on the Earth surface. 
The projects EUSO \cite{EUSO} and OWL \cite{OWL}  
intend to launch 
large aperture optical detectors to 500 km height. The detectors 
would look down upon 
the atmosphere and search for nitrogen fluorescence signals 
due to neutrino interactions.  
The monitored mass would be up to  10 Tera-tons, with an energy threshold 
of about 10$^{10}$ GeV.

\subsection{Scenario for the next decade}

\begin{figure}
\centering
\includegraphics[width=8cm]{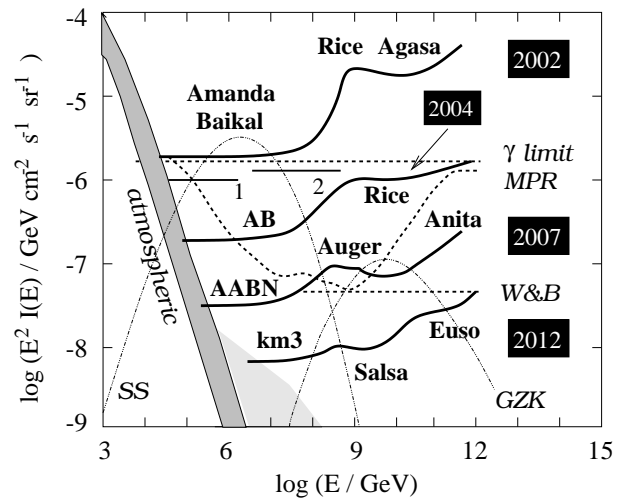}
\caption{\label{diffuse} 
Scenario for the improvement of experimental sensitivities 
to  diffuse extraterrestrial fluxes of high energy neutrinos.
AB = Amanda,\,Baikal, AABN = Amanda,\,Antares,\,Baikal,\,Nestor.
{\it 1,2,:} Amanda limits obtained from the analysis of
upward (1) and high energy downward (2) tracks, assuming
an $E^{-2}$ spectrum. 
The grey band denotes the flux of atmospheric neutrinos, with
the excess at high energies being an estimate for the
contribution from prompt muons and neutrinos due to
charm decays in air showers. 
Dashed lines indicate various theoretical bounds, the 2
thin curves specific flux predictions (see text).}
\end{figure}

\begin{figure}
\centering
\includegraphics[width=7cm]{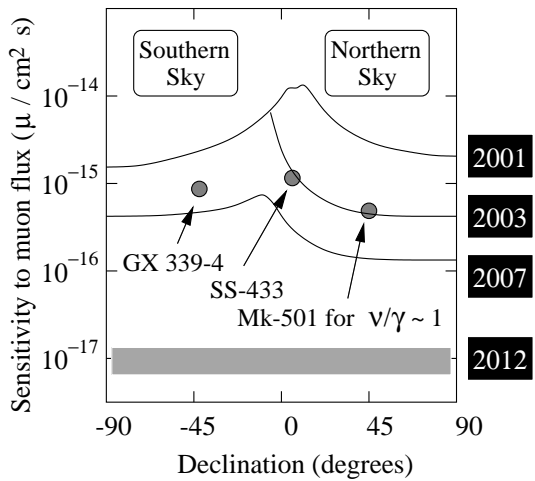}
\caption{\label{point} 
Scenario for the 
improvement of experimental sensitivities to TeV point sources. 
Expected steps for the Northern sky are obtained from Amanda (2003), 
and Amanda together with the first strings of IceCube (2007), 
on the Southern sky from the Mediterranean detectors Antares
and Nestor (2007). 
In 2012, both hemispheres will have profited from cubic 
kilometer arrays indicated by the grey band.
Shown are also predicted fluxes for two microquasars \cite{Guetta} - one in the
northern and one in the southern hemisphere - which are just in
reach for Amanda and the Mediterranean arrays. As a benchmark,
we show also the flux which would be expected if Mk501, a source
spectacular in TeV gamma rays,  would produce a similar
flux in TeV neutrinos.}
\end{figure}

The next ten years promise to be a particularly 
exciting decade for high energy neutrino astrophysics. 
Figures \ref{diffuse} and \ref{point} sketch possible 
scenarios to move the frontier towards unprecedented  sensitivities. 

Figure \ref{diffuse} addresses the sensitivity to diffuse fluxes, 
i.e. integrating over the full angular acceptance of the detectors. 
The scale is set by the known flux of atmospheric neutrinos, 
by the bounds derived from observed fluxes of charged cosmic rays 
(W\&B \cite{WB} and, with less stringent assumptions, 
the lower MPR curve \cite{MPR}), by gamma rays (horizontal
MPR line which assumes that cosmic rays are mostly confined
in the cosmic source region and only gammas and neutrinos escape), 
and by specific model predictions \cite{Semikoz}. The figure shows 
two of the latter, one for the predicted flux of
GZK neutrinos at ultra-high energies (see item {\it b)} above), 
the other for a model of neutrinos from
Active Galactic Nuclei, peaking at TeV-PeV energies (Stecker and
Salomon, SS \cite{SS}).
The majority of the limits shown
are published as ``differential'' limits, defining the
flux sensitivity as the neutrino flux which gives at least one
observed event per decade of energy per year (assuming
negligible background). The limits published for AMANDA
assume an $E^{-2}$ flux. They come from two separate analyses, the
one studying upward muon tracks, the other downward
tracks of very high energy which are unlikely to be due
to muons generated in the atmosphere.
Both analyses, however, properly account for the background of these 
atmospheric muons. 
The lines
marked {\it 1} and {\it 2} extend over the energy range containing
90\% of the events expected from an $E^{-2}$ flux.
For better illustration of the
progress in time and over all the energy range,
the AMANDA limits as well as the limits expected for
ICECUBE have been translated to limits differentially
per energy decade.

The present frontier is defined by TeV-PeV 
limits obtained by AMANDA and BAIKAL, and by PeV-EeV limits from the 
South Pole radio array RICE and the Japanese AGASA air shower array. 
Note that the Baikal/Amanda limits just reached a level
sufficient to test (and actually to exclude) the model shown.
The progress over the next 2 years will come from the same experiments. 
After that, the Mediterranean telescopes -- ANTARES and
NESTOR -- will start to contribute, 
flanked by AUGER and the ANITA balloon mission at high energies. 
This could result in an improvement of about two orders of magnitude 
over the full relevant energy range. Finally, ten years from now, 
the TeV-PeV sensitivity will be dictated by the cubic kilometer 
arrays at the South Pole and in the Mediterranean (marked
as ``km3''). 
At the high energy frontier, a SALSA-like experiment, 
and still higher, satellite detectors, might push the limit down by 
about three orders of magnitude compared to 2002.

Most likely, the first signal with clear signature will be a point source, 
possibly a transient signal which is easiest to identify. 
Figure \ref{point} sketches a possible road until 2012. 
Best present limits are from MACRO, Super-Kamiokande (Southern sky) 
and AMANDA (Northern sky). This picture will not change until 
the Mediterranean detectors come into operation.
AMANDA and ANTARES/NESTOR have the first realistic chance to discover 
an extraterrestrial neutrino source. The ultimate sensitivity for 
the TeV-PeV range is likely reached by the cubic kilometer arrays. 
This scale is set by many model predictions for neutrinos from cosmic 
accelerators or from dark matter decay. However, irrespective of 
any specific model  prediction, these gigantic detectors, 
hundred times larger than AMANDA and thousand times larger than 
underground detectors, will hopefully keep the promise for  
any detector opening  a new window to the Universe: 
to detect {\it unexpected} phenomena.

\end{document}